\documentclass[twocolumn,preprintnumbers,pre]{revtex4}
\usepackage{graphicx, subfigure}
\usepackage{amssymb, amsmath,amssymb,amsfonts}
\usepackage{amsthm,mathrsfs,amsopn}
\usepackage{dcolumn}
\usepackage{bm}
\usepackage{color}
\usepackage[utf8]{inputenc}
\usepackage{mathtools}
\usepackage{nicefrac}

\theoremstyle{plain} %% This is the default

\definecolor{internationalkleinblue}{rgb}{0.0, 0.18, 0.65}
\definecolor{brickred}{rgb}{0.8, 0.25, 0.33}
\definecolor{darkviolet}{rgb}{0.58, 0.0, 0.83}

\begin{document}

\title{Pattern reconstruction through generalized eigenvectors on defective networks}

\author{Marie Dorchain}
\affiliation{Department of Mathematics and naXys, Namur Institute for Complex Systems, University of Namur, Rue Graf\'e 2, B5000 Namur, Belgium}

\author{Riccardo Muolo}
\affiliation{Department of Mathematics and naXys, Namur Institute for Complex Systems, University of Namur, Rue Graf\'e 2, B5000 Namur, Belgium}

\author{Timoteo Carletti}
\email{timoteo.carletti@unamur.be}
\affiliation{Department of Mathematics and naXys, Namur Institute for Complex Systems, University of Namur, Rue Graf\'e 2, B5000 Namur, Belgium}

\begin{abstract}
Self-organization in natural and engineered systems causes the emergence of ordered spatio-temporal motifs. In presence of diffusive species, Turing theory has been widely used to understand the formation of such patterns \textcolor{black}{on continuous domains} obtained from a diffusion-driven instability mechanism. The theory was later extended to networked systems, where the reaction processes occur locally (in the nodes), while diffusion takes place through the networks links. The condition for the instability onset relies on the spectral property of the Laplace matrix, i.e., the diffusive operator, and in particular on the existence of an eigenbasis. In this work we make one step forward and we prove the validity of Turing idea also in the case of a \textcolor{black}{network with} defective Laplace matrix. Moreover, by using both eigenvectors and generalized eigenvectors we show that we can reconstruct the asymptotic pattern with a relatively small discrepancy.
Because a large majority of empirical networks are non-normal and often defective, our results pave the way for a thorough understanding of self-organization in real-world systems.  
\end{abstract}
\maketitle

\noindent \textbf{Introduction.}\,-\,We are surrounded by patterns. Those spatio-temporal motifs are the signature of the emergence of order from disorder~\cite{Anderson} resulting from the collective behavior of the many nonlinearly interacting basic units~\cite{PSV,NicolisPrigogine}. In many relevant applications these interactions can be modeled by using reaction-diffusion equations aiming at describing the behavior of concentrations in time and space, being the latter a continuous domain~\cite{Murray} or a discrete one, e.g., a complex network~\cite{NakaoMikhailov}. Indeed local reactions require, by their very first nature, species to be spatially close, hence separated from other groups; it is thus natural to consider species to occupy spatially limited zones, i.e., nodes of a network, and diffuse across paths connecting different zones, i.e., the links of a network. This will be the framework we will be interested in this work.

Alan Turing introduced and studied in the 50s a symmetry breaking mechanism where a spatially homogeneous equilibrium of a reaction-diffusion system loses its stability once disturbed with an heterogeneous perturbation; eventually the system achieves a new, generally, patchy stationary or oscillatory solution~\cite{Turing}. Nowadays, Turing instability finds application beyond the original framework of morphogenesis or chemical reaction systems~\cite{Castets,DeKepper,Tompkins} and it stands for a pillar to explain self-organization in nature~\cite{NicolisPrigogine,Pismen}, having being formalized by the existence of an interplay between slow diffusing activators and fast diffusing inhibitors~\cite{GiererMeinhardt}. Indeed the latter determines a general mechanism: a local feedback, i.e., short range production of a given species, which should be, at the same time, inhibited at distance, by long range interaction.

The onset of Turing instability ultimately relies on the study of the spectral properties of a suitable operator buil\textcolor{black}{t} by using the Jacobian of the reaction part and the diffusion term, i.e., the Laplace operator. By assuming the existence of an eigenbasis for the latter, one can compute the dispersion relation, that ultimately determines the onset of the instability. Those ideas have been largely applied to study the emergence of Turing patterns for system whose underlying network is symmetric~\cite{NakaoMikhailov}, directed~\cite{Asllani1,carletti}, but also for multiplex~\cite{busiello_turing} and multilayer networks~\cite{Asllani2016}, temporal networks~\cite{PABFC2017} and even in the novel framework of higher-order structures, such as hypergraphs~\cite{muologallo} and simplicial complexes~\cite{turing_topological}. In particular, it has been shown that the final pattern can be partially reconstructed by considering the eigenvectors associated to the unstable modes, namely those for which the dispersion relation is positive. Indeed in the linear regime, namely close to the bifurcation, the pattern is completely aligned with those critical eigenvectors; remarkably enough the nonlinearity of the model only slightly perturbs this behavior and thus the final pattern can be \textcolor{black}{accurately} described by a linear combination of critical eigenvectors~\cite{NakaoMikhailov}. The agreement is stronger the fewer is the number of unstable modes. 

Scholars have recently pointed out that most real-world networks are non-normal~\cite{malbor_teo,DNEM}, namely their adjacency matrix $\mathbf{A}$ does satisfy $\mathbf{A}\mathbf{A}^\top\neq \mathbf{A}^\top\mathbf{A}$~\cite{trefethen}, or equivalently $\mathbf{A}$ is not diagonalizable through an orthonormal transformation. Turing patterns on non-normal networks have been recently studied with a numerical approach~\cite{top_resilience,jtb}. These latter results however still rely on the assumption of the existence of a basis of eigenvectors for the Laplace matrix.

The goal of this work is thus twofold. We first analytically solve the problem on defective networks, i.e., {\textcolor{black}{networks whose Laplace matrix does not admit an eigenbasis, and thus the eigenvalues have algebraic multiplicity larger than one and greater than the geometric multiplicity.}} 
Then, we show how the pattern reconstruction is improved when considering also the generalized eigenvectors associated to the unstable modes.\\

%The work is organized as follows. In the next Section, we will extend the theory for the case of defective networks, building an analytical ground to what has been shown numerically in previous works involving non-normal networks. In Section III, we will numerically show the onset of patterns and show how they are related to some instability modes, associated to the eigenvectors of the Laplace matrix. The main result of the paper will be discussed in Section IV, where we show the role that generalized eigenvectors have in improving the reconstruction process. Finally, Section V is left for discussion and further perspectives.

\noindent\textbf{Turing theory on defective networks.}\,-\,Let us consider two different species populating a directed network composed by $n$ nodes and let us denote by $u_i(t)$ and $v_i(t)$, $i=1,\dots,n$, their respective concentrations on node $i$ at time $t$. When species happen to share the same node, they interact via some generic nonlinear functions $f(u_i,v_i)$ and $g(u_i,v_i)$. On the other hand, they can diffuse across the available network links according to Fick's law by taking into account the link directionality. The model can hence be mathematically cast in the form
\begin{equation}
\label{eq:model}
\begin{dcases}
\frac{du_i}{dt}&=f(u_i,v_i)+D_u\sum_{j=1}^{n}L_{ij} u_j  \\ 
\frac{dv_i}{dt}&=g(u_i,v_i)+D_v\sum_{j=1}^{n}L_{ij} v_j 
\end{dcases} \quad\forall i=1,\dots,n\, ,
\end{equation}
where $D_u>0$ (resp. $D_v>0$) is the diffusion coefficients of species $u$ (resp. $v$). The Laplace matrix, $L_{ij}=A_{ij}-\delta_{ij}k_i^{(in)}$, is the discrete equivalent of the diffusion operator in the continuous support case, where $A_{ij}$ is the $(i,j)$ entry of the adjacency matrix that allows to encode the nodes connections, $A_{ij}=1$ if there is a link pointing from node $j$ to node $i$, and $k_i^{(in)}=\sum_jA_{ij}$ is the in-degree of node $i$.

In the spirit of Turing framework, we assume the existence of a stable solution of~\eqref{eq:model} once we silence the diffusive part, namely there exists $(u_*,v_*)$ such that $f(u_*,v_*)=g(u_*,v_*)=0$ and moreover $\mathrm{tr}(\mathbf{J}_0)<0$ and $\mathrm{det}\mathbf{J}_0>0$, where $\mathbf{J}_0$ is the Jacobian matrix of the reaction part evaluated at the equilibrium $(u_*,v_*)$
\begin{equation}
\label{eq:J0}
 \mathbf{J}_0=\left(
\begin{matrix}
 \partial_u f &  \partial_v f\\
  \partial_u g &  \partial_v g
\end{matrix}
\right)\Big\rvert_{(u_*,v_*)}\, ,
\end{equation}
where we denoted by $\partial_u f$ the derivative of $f$ with respect to $u$ evaluated on the equilibrium $(u_*,v_*)$, and similarly for the other terms.

We then require such equilibrium to turn out unstable once diffusion is at play. To verify such condition \textcolor{black}{we perform a linear stability analysis, namely} we introduce a perturbation from the homogeneous solution $\delta u_i(t)=u_i(t)-u_*$ and $\delta v_i(t)=v_i(t)-v_*$, and expand Eq.~\eqref{eq:model}, keeping only the first order terms in the perturbation (the latter assumed to be small). We thus obtain for all $i=1,\dots,n$
\begin{equation}
\label{eq:model2}
\begin{dcases}
\frac{d\delta u_i}{dt}&=\partial_u f \delta u_i +\partial_v f \delta v_i + D_u\sum_{j=1}^{n}L_{ij} \delta u_j  \\ 
\frac{d\delta v_i}{dt}&=\partial_u g \delta u_i +\partial_v g \delta v_i +D_v\sum_{j=1}^{n}L_{ij} \delta v_j \, ,
\end{dcases}  
\end{equation}
where we employed the fact that $\sum_j L_{ij}=0$ to {nullify the terms $\sum_jL_{ij}u_*$ and $\sum_jL_{ij}v_*$}.

By introducing the $n\times 2$ vector $\delta\mathbf{x}=(\delta u_1,\delta v_1,\dots,\delta u_n,\delta v_n)^\top$, we can eventually rewrite the latter equation in a compact form as:
\begin{equation}
 \frac{d\delta\mathbf{x}}{dt}=\left[ \mathbf{I}_n\otimes \mathbf{J}_0+ \mathbf{L} \otimes \mathbf{D}\right]\delta\mathbf{x}\, ,
\label{eq:lincompact}
\end{equation}
where $\otimes$ denotes the Kronecker product of matrices, $\mathbf{I}_n$ is the $n \times n$ identity matrix and $\mathbf{D}=\left(
\begin{smallmatrix}
 D_u& 0\\0 & D_v
\end{smallmatrix}
\right)$.

To make some analytical progress, the standard step is thus to simplify the previous $2n\times 2n$ system into $n$ systems $2\times 2$ by assuming the existence of an eigenbasis for the Laplace matrix and projecting the perturbations $\delta u_i$ and $\delta v_i$ upon such basis. Our goal is to show that one can obtain a similar understanding of the onset of Turing instability also in the case the Laplace matrix is defective. %, namely it does not exhibit an eigenbasis. 
Such framework has been studied in~\cite{mott_nish2} in the study of synchronization of coupled oscillators. 

\textcolor{black}{To achieve this goal,} we can invoke the Jordan canonical form to determine an invertible $n\times n$ matrix $\mathbf{P}$ such that
\begin{equation*}
\mathbf{P}^{-1}\mathbf{L}\mathbf{P}=\mathbf{B}=\mathrm{diag}(\mathbf{B}_1,\dots,\mathbf{B}_\ell)\, ,
\end{equation*}
where the $\mathbf{B}_j$ is the $m_j\times m_j$ Jordan block, \textcolor{black}{being $m_j$ the algebraic multiplicity of the eigenvalue $\Lambda^{(j)}$,} $m_1+\dots+m_\ell=n$ and
\begin{equation}
\label{eq:Blambdaj}
\mathbf{B}_j=\left(
\begin{matrix}
 \Lambda^{(j)} & & & \\
 1&  \Lambda^{(j)}&  & \\
  &   \ddots& \ddots &\\
   & & 1& \Lambda^{(j)}
\end{matrix}
\right)\, .
\end{equation}

Let us consider again Eq.~\eqref{eq:lincompact}. By defining $\mathbf{Q}=\mathbf{P}\otimes \mathbf{I}_2$ and ${\color{black}{\delta\mathbf{y}=\mathbf{Q}^{-1}\delta\mathbf{x}}}$ we get
\begin{eqnarray}
 \frac{d\delta\mathbf{y}}{dt} &=&\mathbf{Q}^{-1}\frac{d\delta\mathbf{x}}{dt}\notag \\&=&(\mathbf{P}^{-1}\otimes \mathbf{I}_2) \left[ \mathbf{I}_n\otimes\mathbf{J}_0+ \mathbf{L} \otimes \mathbf{D}\right](\mathbf{P}\otimes \mathbf{I}_2)\mathbf{Q}^{-1}\delta\mathbf{x} \notag \\&=&\left[ \mathbf{I}_n\otimes\mathbf{J}_0+\mathbf{B} \otimes \mathbf{D}\right]\delta\mathbf{y}\, .
\label{eq:lincompactJordan}
\end{eqnarray} 
The vector $\delta\mathbf{y}$ inherits the Jordan decomposition, hence we can write $\delta\mathbf{y}=((\delta\mathbf{y}^{(1)})^\top,\dots,(\delta\mathbf{y}^{(\ell)})^\top)^\top$, where $\delta\mathbf{y}^{(j)}$ is a $(2\times m_j)$-dimensional vector. \\

The stability properties of $\delta \mathbf{y}$ will thus be determined by analyzing the behavior of $\delta \mathbf{y}^{(j)}$, $j=1,\dots,\ell$. Let us consider separately the case $\Lambda^{(1)}=0$ and ${\color{black}{\Re\Lambda^{(j)}<0}}$ for $j\geq 2$. Assume thus $\Lambda^{(1)}=0$ to be degenerate and be $m_1$ its multiplicity. If $m_1=1$ then $\delta \mathbf{y}^{(1)}$ evolves accordingly to $\mathbf{I}_n\otimes \mathbf{J}_0$ and thus $\delta \mathbf{y}^{(1)}$ is stable because of the condition imposed on the homogeneous equilibrium $(u_*,v_*)$~\footnote{Let us observe that a similar result can be obtained once the degeneracy is a consequence of the presence of nodes without incoming links, called \textit{leader nodes}~\cite{asllani_leaders}, as the corresponding eigenvectors are the canonical ones, and the degeneration is algebraic and not geometric.}. Otherwise $\mathbf{B}_1$ is a $m_1\times m_1$ matrix of the form~\eqref{eq:Blambdaj} with $0$ on the diagonal.

The part of Eq.~\eqref{eq:lincompactJordan} relative to $\Lambda^{(1)}$ can thus be rewritten as
\begin{eqnarray}
 \frac{d\delta\mathbf{y}^{(1)}}{dt} &=&\left[\left(
\begin{matrix}
 \mathbf{J}_0 & &\\
 & \ddots &\\
 & &  \mathbf{J}_0
\end{matrix}\right)+ 
\left(
\begin{matrix}
 0& & & \\
 \mathbf{D}&  0&  & \\
  &   \ddots& \ddots &\\
   & & \mathbf{D}& 0
\end{matrix}
\right)\right]\delta\mathbf{y}^{(1)} \nonumber \, ,
\label{eq:lincompactJordan1}
\end{eqnarray}
the matrix on the right hand side has the same eigenvalues of $\mathbf{J}_0$ from which we can conclude that $\delta\mathbf{y}^{(1)}$ is stable.
% \begin{eqnarray}
%  \frac{d\delta\mathbf{y}^{(1)}}{dt} &=&\left(
% \begin{matrix}
%  \mathbf{J}_0 & &\\
%  & \ddots &\\
%  & &  \mathbf{J}_0
% \end{matrix}\right)\delta\mathbf{y}^{(1)}+  \\
% & &+\left(
% \begin{matrix}
%  0& & & \\
%  \mathbf{D}&  0&  & \\
%   &   \ddots& \ddots &\\
%    & & \mathbf{D}& 0
% \end{matrix}
% \right)\delta\mathbf{y}^{(1)} \nonumber \, .
% \label{eq:lincompactJordan1}
% \end{eqnarray}

%Let us now consider the latter equation by writing $\delta\mathbf{y}^{(1)}=(\xi^{(1)}_1,\eta^{(1)}_1,\dots,\xi^{(1)}_{m_1},\eta^{(j)}_{m_1})^\top$, where $(\xi^{(1)}_i,\eta^{(1)}_i)\in\mathbf{R}^2$ for all $i=1,\dots,m_1$. We get
%\begin{eqnarray}
%\frac{d}{dt}\binom{\xi^{(1)}_1}{\eta^{(1)}_1} &=&\mathbf{J}_0\binom{\xi^{(1)}_1}{\eta^{(1)}_1} \label{eq:lincompactJordanj1_lambda1}\, , \\
%\frac{d}{dt}\binom{\xi^{(1)}_2}{\eta^{(1)}_2} &=&\mathbf{J}_0\binom{\xi^{(1)}_2}{\eta^{(1)}_2}+\mathbf{D}\binom{\xi^{(1)}_1}{\eta^{(1)}_1}\, , \notag \\
%& \vdots &  \label{eq:lincompactJordanjmj_lambda1}\\
%\frac{d}{dt}\binom{\xi^{(1)}_{m_1}}{\eta^{(1)}_{m_1}} &=&\mathbf{J}_0\binom{\xi^{(1)}_{m_1}}{\eta^{(1)}_{m_1}}+\mathbf{D}\binom{\xi^{(1)}_{m_1-1}}{\eta^{(1)}_{m_1-1}} \notag \, .
%\end{eqnarray}

%The first Eq.~\eqref{eq:lincompactJordanj1_lambda1} is the equation for the isolated system, which is set to be stable. By considering Eqs.~\eqref{eq:lincompactJordanjmj_lambda1}, it can be proven by recurrence that the system remains stable, given that the first part is again the isolated system, while the second is also stable because solution of the previous (stable) system.  \\

We can now analyze the remaining cases $j=2,\dots, \ell$. The part of Eq.~\eqref{eq:lincompactJordan} involving $\delta\mathbf{y}^{(j)}$ is thus
\begin{eqnarray}
 \frac{d\delta\mathbf{y}^{(j)}}{dt} &=&\left(
\begin{matrix}
 \mathbf{J}_0 & &\\
 & \ddots &\\
 & &  \mathbf{J}_0
\end{matrix}\right)\delta\mathbf{y}^{(j)}+  \\
& &+\left(
\begin{matrix}
 \Lambda^{(j)} \mathbf{D}& & & \\
 \mathbf{D}&  \Lambda^{(j)}\mathbf{D}&  & \\
  &   \ddots& \ddots &\\
   & & \mathbf{D}& \Lambda^{(j)}\mathbf{D}
\end{matrix}
\right)\delta\mathbf{y}^{(j)} \nonumber \, .
\label{eq:lincompactJordanj}
\end{eqnarray}
We can reformulate the previous equation by writing $\delta\mathbf{y}^{(j)}=(\xi^{(j)}_1,\eta^{(j)}_1,\dots,\xi^{(j)}_{m_j},\eta^{(j)}_{m_j})^\top$, where {\color{black}{$\delta\mathbf{y}_i^{(j)}$}}$=(\xi^{(j)}_i,\eta^{(j)}_i)\in\mathbb{R}^2$ for all $i=1,\dots,m_j$, thus obtaining
\begin{eqnarray}
  \frac{d}{dt}\binom{\xi^{(j)}_1}{\eta^{(j)}_1} &=&\mathbf{J}_0\binom{\xi^{(j)}_1}{\eta^{(j)}_1}+\Lambda^{(j)} \mathbf{D}\binom{\xi^{(j)}_1}{\eta^{(j)}_1}
 \label{eq:lincompactJordanj1}\, , \\
  \frac{d}{dt}\binom{\xi^{(j)}_2}{\eta^{(j)}_2} &=&\mathbf{J}_0\binom{\xi^{(j)}_2}{\eta^{(j)}_2}+\Lambda^{(j)} \mathbf{D}\binom{\xi^{(j)}_2}{\eta^{(j)}_2}+\mathbf{D}\binom{\xi^{(j)}_1}{\eta^{(j)}_1}\, , \notag \\
& \vdots & \label{eq:lincompactJordanjmj}\\
 \frac{d}{dt}\binom{\xi^{(j)}_{m_j}}{\eta^{(j)}_{m_j}} &=&\mathbf{J}_0\binom{\xi^{(j)}_{m_j}}{\eta^{(j)}_{m_j}}+\Lambda^{(j)} \mathbf{D}\binom{\xi^{(j)}_{m_j}}{\eta^{(j)}_{m_j}}+\mathbf{D}\binom{\xi^{(j)}_{m_j-1}}{\eta^{(j)}_{m_j-1}}  \, \notag.
\end{eqnarray}

The first Eq.~\eqref{eq:lincompactJordanj1} is the same equation one would get once the Laplace matrix admits an eigenbasis, thus one can determine a condition on $\Lambda^{(j)}$ to make the projection $\delta\mathbf{y}_1^{(j)}=(\xi^{(j)}_1,\eta^{(j)}_1)^\top$ unstable~\cite{Asllani1}, namely to compute the eigenvalue with the largest real part of the matrix $\mathbf{M}_j=\mathbf{J}_0+\Lambda^{(j)} \mathbf{D}$ (see also Appendix~\ref{sec:instreg}).

Let us now consider Eqs.~\eqref{eq:lincompactJordanjmj} and observe that each of them is composed by two terms, the first one involving the same matrix of the former equation, $\mathbf{M}_j$, while the second one depends on the projection $\delta\mathbf{y}^{(j)}(t)$. If, for the choice of $\Lambda^{(j)}$ the matrix $\mathbf{M}_j$ is unstable\textcolor{black}{, namely its spectrum contains eigenvalues with positive real part,} and thus $\delta\mathbf{y}_1^{(j)}(t)$ has an exponential growth, then the same is true for $\delta\mathbf{y}_2^{(j)}(t)=(\xi^{(j)}_2,\eta^{(j)}_2)^\top$. By considering the remaining equations and by exploiting the peculiar lower triangular shape of the system, we can prove that if Eq.~\eqref{eq:lincompactJordanj1} returns an unstable solution, then all the solutions $\delta\mathbf{y}_i^{(j)}(t)=(\xi^{(j)}_i,\eta^{(j)}_i)^\top$ are unstable as well.

In conclusion one can compute the {\em dispersion relation}, $\lambda(\zeta)$, namely \textcolor{black}{to determine} the largest real part of the eigenvalues of the $1$-(complex) parameter family system $\mathbf{J}_0+\zeta \mathbf{D}$; if the Laplace matrix $\mathbf{L}$ is defective one can check the instability condition on the available eigenvalues. This accounts to study the sign of $\lambda(\Lambda^{(j)})$, and conclude about the emergence of patterns solely based on this information. Let us observe that this is a sufficient condition, indeed it can happen that the matrix $\mathbf{M}_j$ is stable for all $j$, {\color{black}{i.e., all its eigenvalues have negative real part}}, but the presence of Jordan blocks introduces a transient (polynomial) growth in the linear regime that results strong enough to limit the validity of the linear approximation\textcolor{black}{. Stated differently, the size of the basin of attraction of the stable fixed point considerably shrinks because of this transient growth}. Thus the nonlinear system could exhibit orbits departing from the homogeneous reference solution; only infinitesimal perturbations will be attracted to the latter, %. Stated differently, the stability basin of the reference solution shrinks considerably in presence of defective Laplace matrix, 
the solution is thus stable but finite perturbations can be amplified. Hence the latter result extends and completes \textcolor{black}{the numerical analysis performed in the case of} diagonalizable non-normal networks~\cite{jtb,entropy}.\\

\noindent \textbf{A case study: the Brusselator model.}\,-\,Let us present the described theory by considering the Brusselator model~{\cite{PrigogineNicolis1967,PrigogineLefever1968,galla}}, often invoked in the literature as {a paradigm} nonlinear reaction scheme for studying self-organized phenomena such as synchronization~\cite{entropy}, Turing patterns~\cite{Asllani1} and oscillation death~\cite{KVK2013,LFCP2018}. The key feature of the model is the presence of two species, reacting via a cubic nonlinearity
\begin{equation}
\label{eq:bxl}
\begin{dcases}
\frac{d u}{dt} &= 1-(b+1)u+c u^2v\\
\frac{d v}{dt} &= bu-c u^2v\, ,
\end{dcases}
\end{equation}
where $b>0$ and $c>0$ act as tunable model parameters. One can easily realize the existence of a unique equilibrium $u_* = 1$ and $v_* = b/c$, that results stable if the Jacobian of the reaction part evaluated on it, $\mathbf{J}_{0}=\left( 
\begin{smallmatrix}
b-1 & c \\-b & -c 
\end{smallmatrix}\right)$, has a negative trace, $\mathrm{tr}(\mathbf{J}_{0})=b-c<1$, and a positive determinant $\det \mathbf{J}_{0}=c>0$.

\begin{figure}[ht!]
    \centering
    \includegraphics[width=0.9\linewidth]{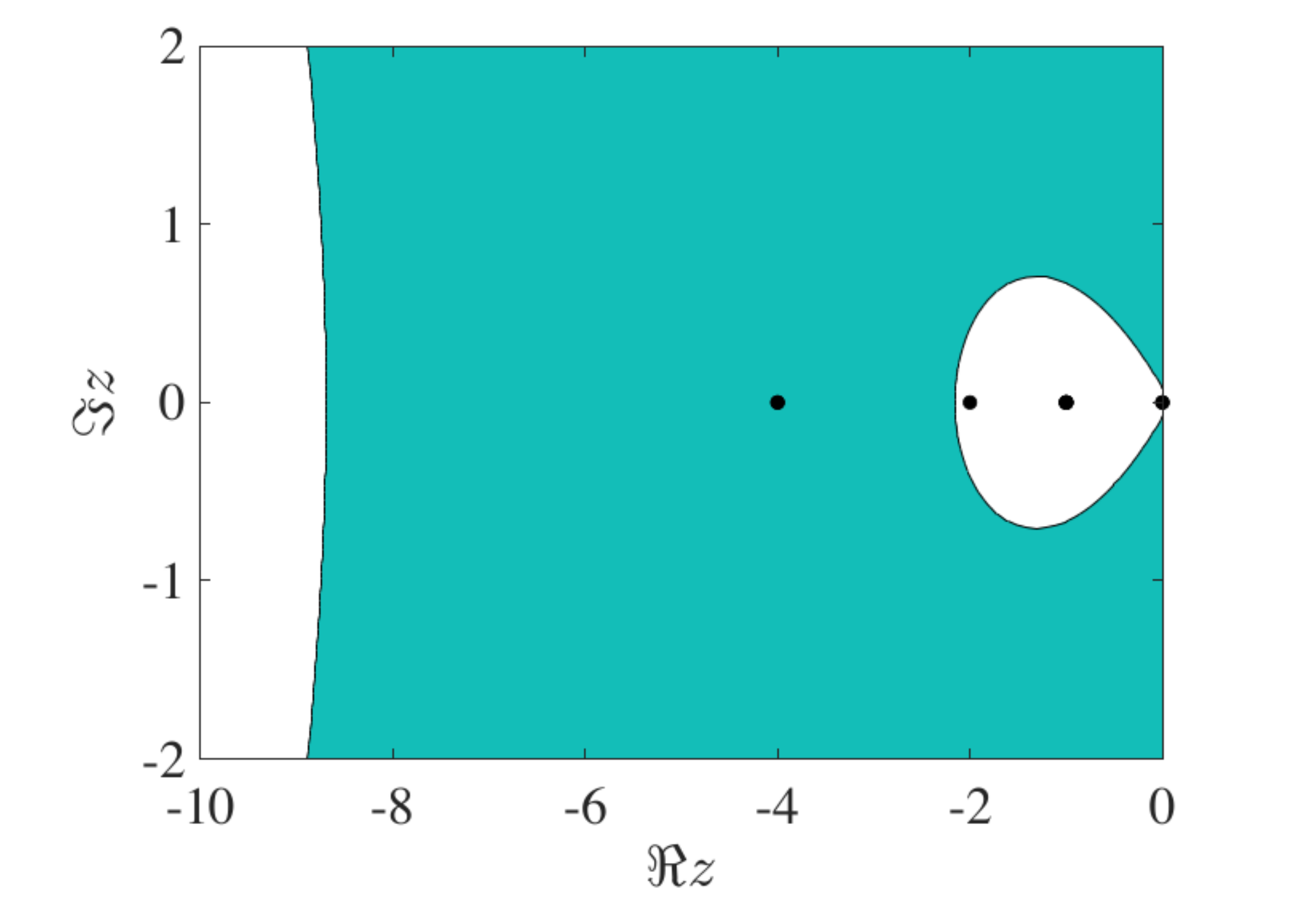}\\
     \includegraphics[width=0.9\linewidth]{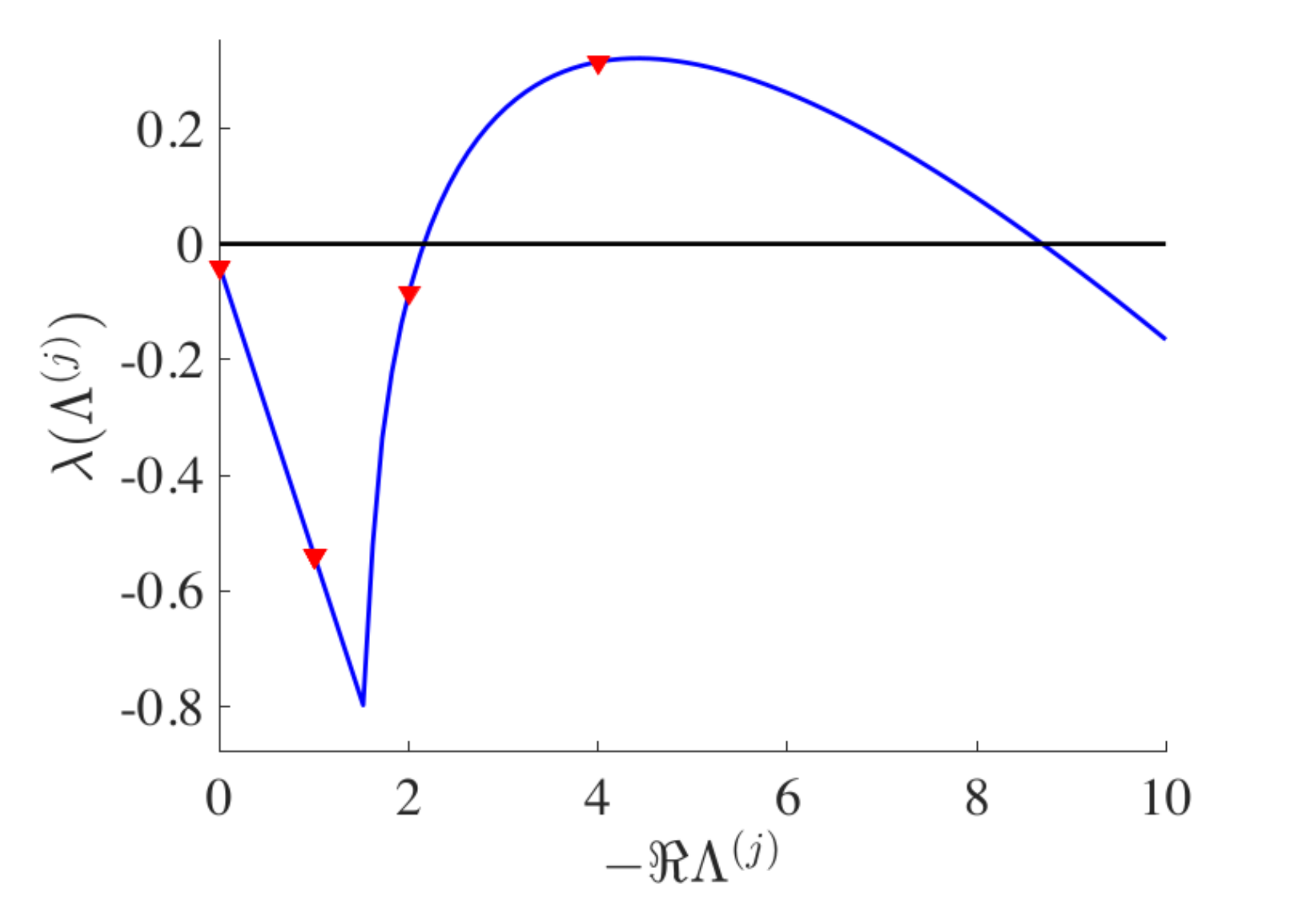}
    \caption{Region of the complex plane associated to Turing instability (top panel) and dispersion relation (bottom panel) computed for the Brusselator model with parameters $b=3.92$, $c=3$, $D_u=0.2$ and $D_v=0.8$. \textcolor{black}{Top panel:} The black dots denote the eigenvalues of the Laplace matrix, $\Lambda^{(1)}=0$ (multiplicity $2$), $\Lambda^{(2)}=-1$ (multiplicity $5$), $\Lambda^{(3)}=-2$ (multiplicity $1$), $\Lambda^{(4)}=-4$ (multiplicity $2$), \textcolor{black}{the green region is associated to a positive dispersion relation, while the white one to the negative case. Bottom panel: the largest real part of the spectrum of the matrix $\mathbf{J}_0+\zeta \mathbf{D}$ is shown in blue as a function of $\zeta$, the dispersion relation evaluated on the Laplace spectrum is reported by using red triangles}.}
    \label{fig:Reginst}
\end{figure}

\begin{figure}[ht!]
    \centering
\includegraphics[width=0.9\linewidth]{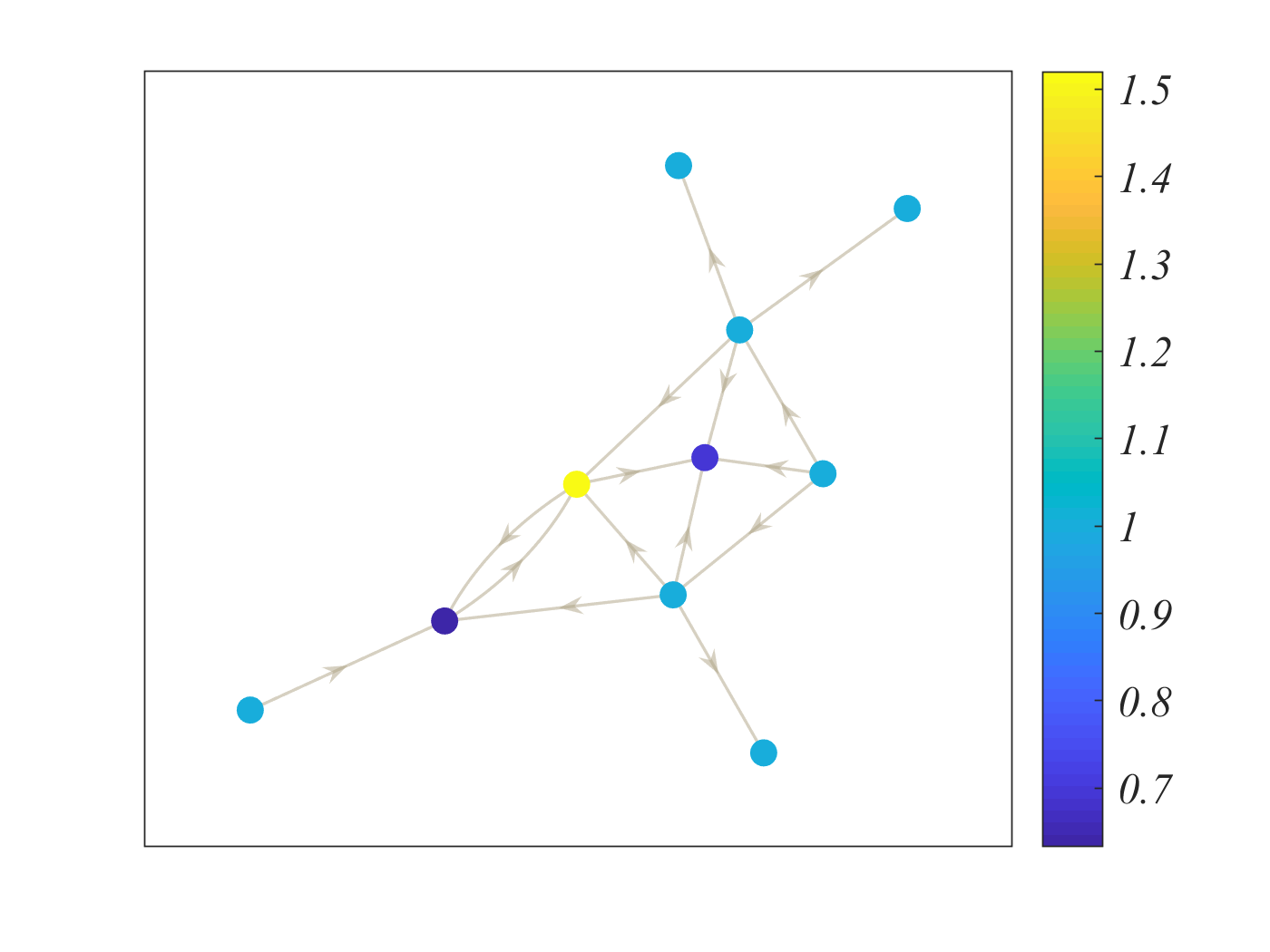}
    \caption{Random non-normal defective network composed by $n=10$ nodes, built by using a directed Erd\H{o}s-R\'enyi \textcolor{black}{algorithm} where the probability to create a bidirectional link is $0.2$ and the probability to transform it into a directed one is $0.6$. \textcolor{black}{Nodes have been colored according to the value of species $u$ at time $\hat{t}=200$ (see colorbar).}}
    \label{fig:3 unstable eigenvalue - network}
\end{figure}

By considering $n$ identical copies of the Brusselator model, each one anchored on a node of a network and interacting with the first neighbors, we obtain $\forall i=1,\dots, n$
\begin{equation}
 \label{eq:sysnABxl}
 \begin{dcases}
\frac{d u_i}{dt} &= 1-(b+1)u_i+c u_i^2v_i+D_u\sum_{j=1}^n {L}_{ij}u_j\\
\frac{d v_i}{dt} &= bu_i-c u_i^2v_i+D_v\sum_{j=1}^n {L}_{ij}v_j \, .
\end{dcases}  
\end{equation}
By linearizing the above equation about the homogeneous equilibrium and by using the Jordan blocks we can obtain the analogous of Eqs.~\eqref{eq:lincompactJordanj1} -~\eqref{eq:lincompactJordanjmj}. Then, we can determine the region in the complex plane (see Appendix~\ref{sec:instreg}) associated to the Turing instability if at least one eigenvalue of the Laplace matrix falls into this region. {\color{black}Let us remark that in the following we always deal with an unstable dispersion relation, i.e., there exists at least one eigenvalue $\Lambda^{(j)}$ such that Eq. \eqref{eq:lincompactJordanj} is unstable.} In the top panel of Fig.~\ref{fig:Reginst} we show the region of instability for the Brusselator model defined on a defective network composed by $N=10$ nodes (see Fig.~\ref{fig:3 unstable eigenvalue - network}), where the stable eigenvalues are $\Lambda^{(1)}=0$, with multiplicity $2$, $\Lambda^{(2)}=-1$ with multiplicity $5$ and $\Lambda^{(3)}=-2$ with multiplicity $1$. There is only one unstable eigenvalue $\Lambda^{(4)}=-4$ with multiplicity $2$.

\begin{figure}[ht!]
    \centering
\includegraphics[width=0.9\linewidth]{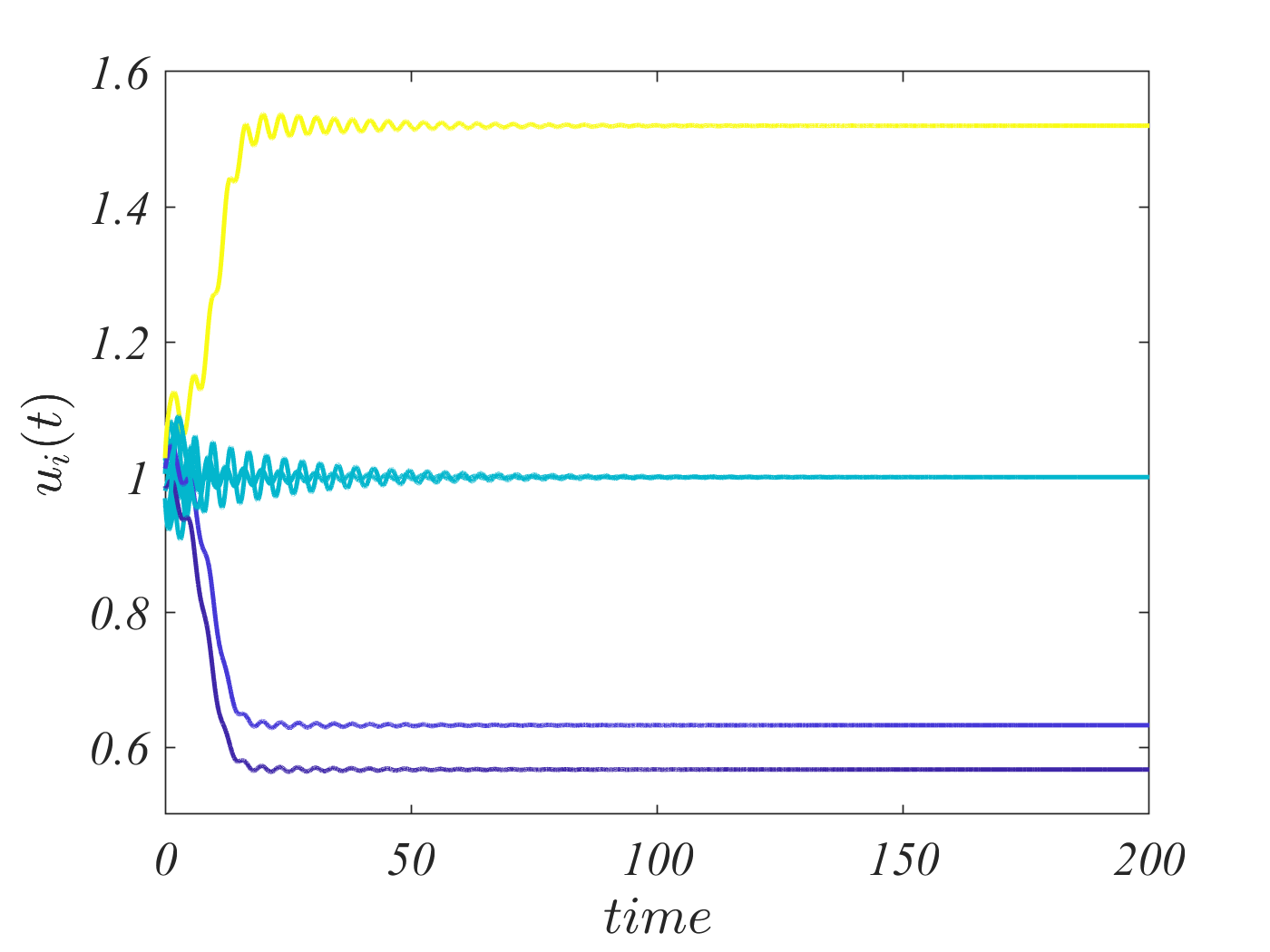}
    \caption{Evolution of the concentration of the specie $u_i$ over time for the Brusselator model with parameters $b=3.92$, $c=3$, $D_u=0.2$ and $D_v=0.8$. The underlying network is the one shown in Fig.~\ref{fig:3 unstable eigenvalue - network}. The nodes concentrations have been initialized to the homogeneous equilibrium, $u_*=1$, upon which a random, node dependent, perturbation of order $0.01$ has been added. The resulting orbits have been obtained by using a Runge-Kutta 4th method with time step $0.01$. \textcolor{black}{Each trajectory has been represented by using the same color of the corresponding node in Fig.~\ref{fig:3 unstable eigenvalue - network}, namely the value at time  $\hat{t}=200$.}}
    \label{fig:3 unstable eigenvalue - pattern}
\end{figure}

In the bottom panel of Fig.~\ref{fig:Reginst} we report the dispersion relation computed for the Brusselator defined on the same defective network. The Laplace eigenvalues are represented by symbols (red triangles) while the continuous curve is the dispersion relation computed for the $1$-(complex) parameter family of linear systems introduced above with the matrix $\mathbf{M}_j$. One can observe that $\lambda(\Lambda^{(j)})$ is positive for $\Lambda^{(4)}$ and thus the equilibrium $(u_*,v_*)\sim (1,1.3067)$ of the coupled system is unstable, as we can appreciate by inspecting Fig.~\ref{fig:3 unstable eigenvalue - pattern} where we report the time evolution of the concentrations $u_i(t)$ vs. time. The same conclusion can be obtained by observing the top panel of Fig.~\ref{fig:Reginst} where we can realize that $\Lambda^{(4)}$ lies inside the instability region \textcolor{black}{(green area)}.\\

\bigskip

\noindent\textbf{Pattern reconstruction through generalized eigenvectors.}\, -\, When there exists an eigenbasis for the Laplace matrix, we can show that pattern can be described by using the eigenvectors related to the unstable eigenvalues, i.e., the eigenvalues of the Laplace matrix that return a positive dispersion relation or equivalently they lie in the instability region as shown above. Our goal is to show that a similar result can be obtained in the case of defective Laplace matrix by recurring to generalized eigenvectors to reconstruct the pattern.
To achieve such goal, we use again the Brusselator model~\eqref{eq:sysnABxl} defined on top of the defective random non-normal network presented above (see Fig.~\ref{fig:3 unstable eigenvalue - network}). Let us remember that the Laplace matrix of the network has an unstable eigenvalue, $\Lambda^{(4)}$ with multiplicity $2$, to which we associate the eigenvector $\phi^{(4)}$ and a generalized eigenvector $v^{(4)}$. 

\begin{figure}[ht!]
    \centering
\includegraphics[width=\linewidth]{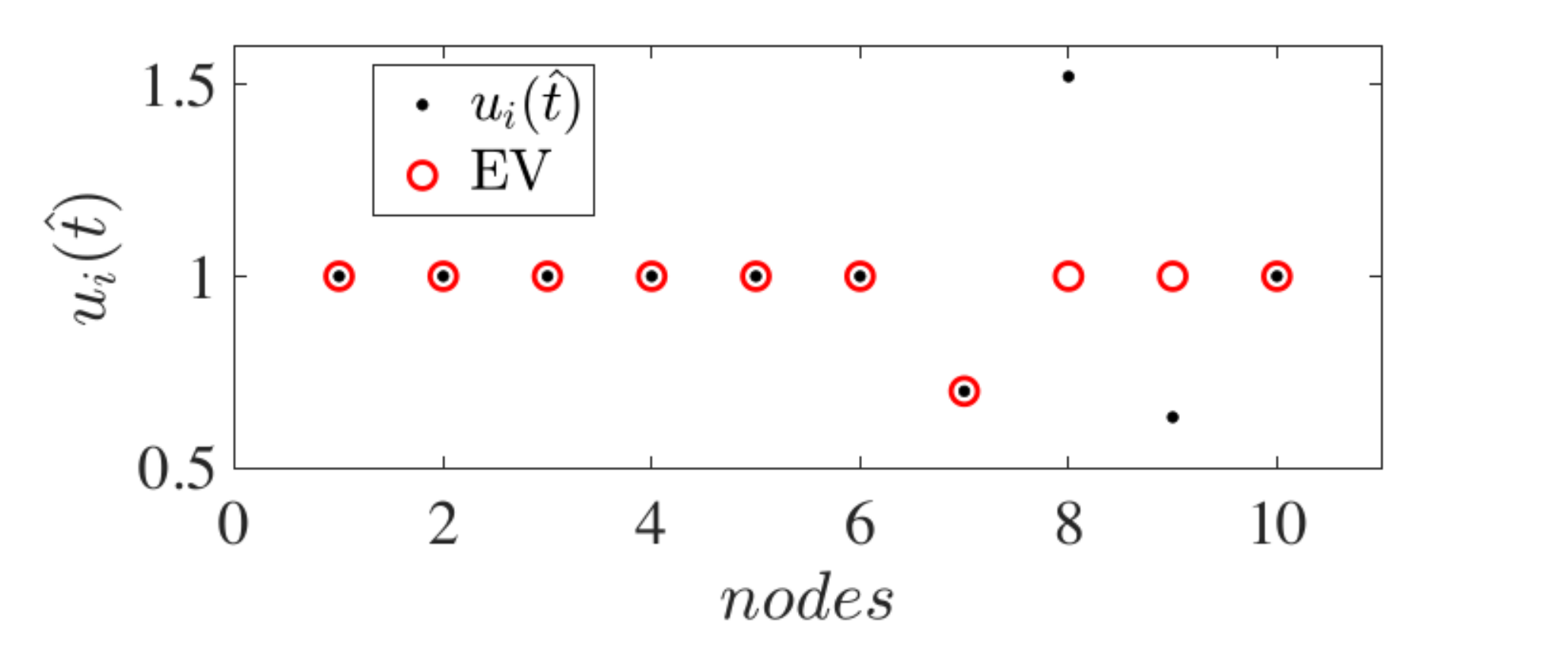}\\
\includegraphics[width=\linewidth]{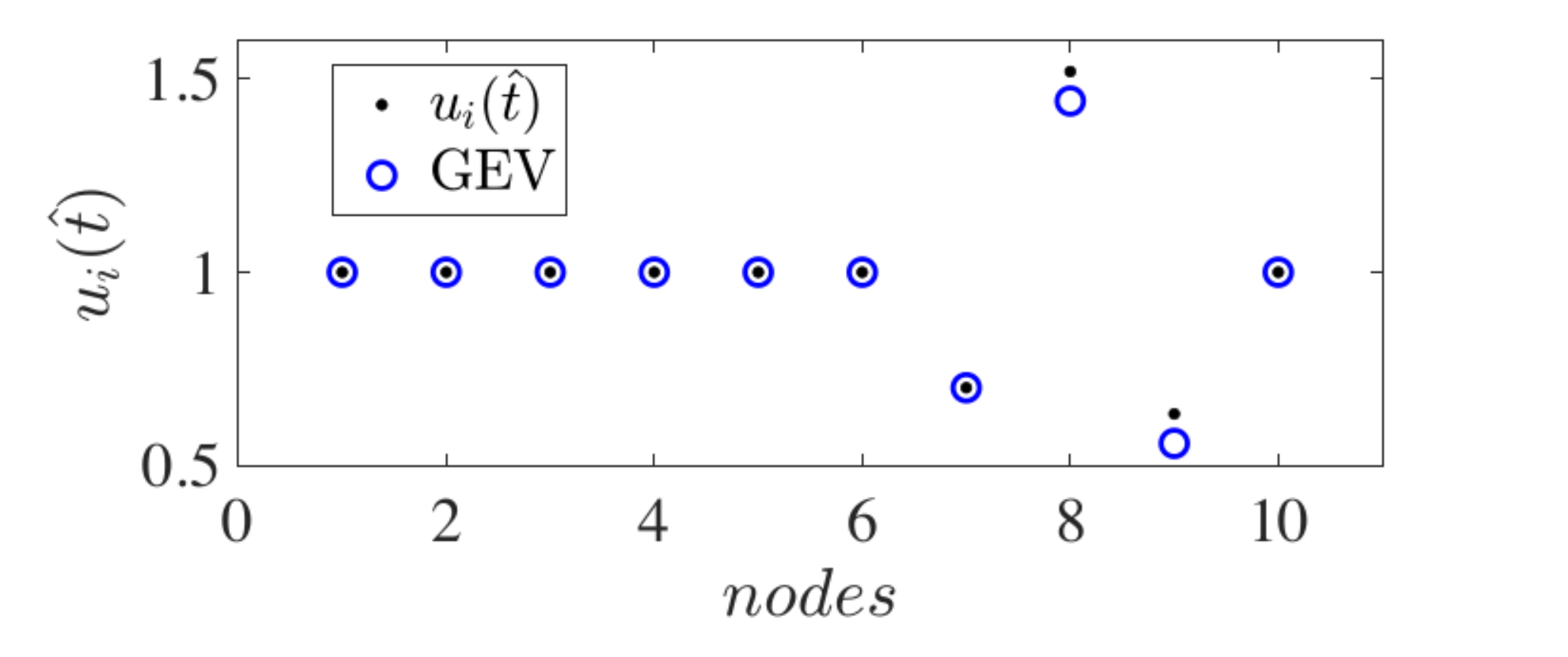}
    \caption{Pattern vs. reconstructed pattern for the Brusselator model with parameters $b=3.92$, $c=3$, $D_u=0.2$ and $D_v=0.8$. Upper panel: the reconstruction is obtained with only the eigenvectors (EV). Lower panel: the eigenvectors and generalized eigenvectors (GEV) are used for the reconstruction. In the first case, the reconstruction error is $\varepsilon=0.088$ while in the second one we have $\varepsilon=0.015$.}
    \label{fig:3unsteigrec}
\end{figure}

To support our claim, we considered the solution $u_i(t)$ of model~\eqref{eq:sysnABxl} up to certain (large) time, $\hat{t}=250$, and we thus obtain the vector $\hat{u}=(u_1(\hat{t}),\dots,u_{10}(\hat{t}))^\top$, i.e., the asymptotic pattern. We then proceed by reconstructing (see Appendix~\ref{sec:recontr}) such pattern by using the eigenvector $\phi^{(4)}$ (see upper panel of Fig.~\ref{fig:3unsteigrec}) or the above eigenvector together with the generalized eigenvector $v^{(4)}$ (see lower panel of Fig.~\ref{fig:3unsteigrec}). We can observe that in both cases the reconstructed pattern is very close to the original one and moreover the one obtained by using also the generalized eigenvector is noticeably improved, having a smaller error. Let us observe that as the reconstruction would obviously be better with two vectors than with one, a weighted absolute error was used to compare the accuracy of the reconstruction. Indeed Each absolute error is weighted by the number of (eigen)vectors used, divided by the total number of \textcolor{black}{available} unstable (eigen)vectors, \textcolor{black}{namely the geometric multiplicity. In this case we hence obtain} $\nicefrac{1}{2}$ in the first case and $\nicefrac{2}{2}$ in the second one (see Appendix~\ref{sec:recontr} \textcolor{black}{for a rigorous definition of the used weighted error}).
Let us notice that our results are robust with respect to the number of (generalized) unstable eigenvectors \textcolor{black}{and their multiplicity (see Appendix~\ref{eq:otherex} and Fig.~\ref{fig:errorvsm})}, the chosen value of $\hat{t}$ or, in the case of oscillatory pattern, to the fact of considering the time-average of $u_i(t)$, i.e., $\langle u \rangle=(\langle u_1 \rangle,\dots,\langle u_{10} \rangle)^\top$ where $\langle u \rangle_i$ is the time-average of the orbit of the $i$-th node (see Appendix~\ref{eq:otherex} and Fig.~\ref{fig:3 unstable eigenvalue - mean pattern reconstruction}). \textcolor{black}{Let us conclude by observing that the proposed pattern reconstruction method scales well with the increasing size of the network. Moreover, the results we obtain by relying on a family of directed defective networks, support the claim that the use of generalized eigenvectors always returns a better estimate of the pattern, than the eigenvectors alone (see Appendix~\ref{eq:otherex} and Fig.~\ref{fig:errorvsN}).}

\noindent\textbf{Conclusions.}\,-\, In this paper we have further extended the Turing theory of pattern formation by studying the case of defective networks. After solving the problem analytically and showing the effects on the instability mechanism given by the presence of Jordan blocks, we have shown the pivotal role of generalized eigenvectors in the reconstruction of the asymptotic pattern. Considering that most real-world networks are non-normal, our results become particularly relevant even looking at other nonlinear phenomena beyond Turing pattern formation and may help in filling the gap between theory and observations. Moreover, the proposed pattern reconstruction method further improves the understanding of the interplay between the dynamics and the underlying topology, paving the way for finer methods of network reconstruction from observational data.

\paragraph*{Acknowledgements} R.M. acknowledges funding from the FNRS, Grant FC 33443, funded by the Walloon region.

\bibliographystyle{apsrev4-1}
\bibliography{biblio_RTP}

%merlin.mbs apsrev4-1.bst 2010-07-25 4.21a (PWD, AO, DPC) hacked
%Control: key (0)
%Control: author (72) initials jnrlst
%Control: editor formatted (1) identically to author
%Control: production of article title (-1) disabled
%Control: page (0) single
%Control: year (1) truncated
%Control: production of eprint (0) enabled
\begin{thebibliography}{33}%
\makeatletter
\providecommand \@ifxundefined [1]{%
 \@ifx{#1\undefined}
}%
\providecommand \@ifnum [1]{%
 \ifnum #1\expandafter \@firstoftwo
 \else \expandafter \@secondoftwo
 \fi
}%
\providecommand \@ifx [1]{%
 \ifx #1\expandafter \@firstoftwo
 \else \expandafter \@secondoftwo
 \fi
}%
\providecommand \natexlab [1]{#1}%
\providecommand \enquote  [1]{``#1''}%
\providecommand \bibnamefont  [1]{#1}%
\providecommand \bibfnamefont [1]{#1}%
\providecommand \citenamefont [1]{#1}%
\providecommand \href@noop [0]{\@secondoftwo}%
\providecommand \href [0]{\begingroup \@sanitize@url \@href}%
\providecommand \@href[1]{\@@startlink{#1}\@@href}%
\providecommand \@@href[1]{\endgroup#1\@@endlink}%
\providecommand \@sanitize@url [0]{\catcode `\\12\catcode `\$12\catcode
  `\&12\catcode `\#12\catcode `\^12\catcode `\_12\catcode `\%12\relax}%
\providecommand \@@startlink[1]{}%
\providecommand \@@endlink[0]{}%
\providecommand \url  [0]{\begingroup\@sanitize@url \@url }%
\providecommand \@url [1]{\endgroup\@href {#1}{\urlprefix }}%
\providecommand \urlprefix  [0]{URL }%
\providecommand \Eprint [0]{\href }%
\providecommand \doibase [0]{http://dx.doi.org/}%
\providecommand \selectlanguage [0]{\@gobble}%
\providecommand \bibinfo  [0]{\@secondoftwo}%
\providecommand \bibfield  [0]{\@secondoftwo}%
\providecommand \translation [1]{[#1]}%
\providecommand \BibitemOpen [0]{}%
\providecommand \bibitemStop [0]{}%
\providecommand \bibitemNoStop [0]{.\EOS\space}%
\providecommand \EOS [0]{\spacefactor3000\relax}%
\providecommand \BibitemShut  [1]{\csname bibitem#1\endcsname}%
\let\auto@bib@innerbib\@empty
%</preamble>
\bibitem [{\citenamefont {Anderson}(1972)}]{Anderson}%
  \BibitemOpen
  \bibfield  {author} {\bibinfo {author} {\bibfnamefont {P.}~\bibnamefont
  {Anderson}},\ }\href@noop {} {\bibfield  {journal} {\bibinfo  {journal}
  {Phys. Rev. Lett.}\ }\textbf {\bibinfo {volume} {177}} (\bibinfo {year}
  {1972})}\BibitemShut {NoStop}%
\bibitem [{\citenamefont {Pastor-Satorras}\ and\ \citenamefont
  {Vespignani}(2010)}]{PSV}%
  \BibitemOpen
  \bibfield  {author} {\bibinfo {author} {\bibfnamefont {R.}~\bibnamefont
  {Pastor-Satorras}}\ and\ \bibinfo {author} {\bibfnamefont {A.}~\bibnamefont
  {Vespignani}},\ }\href@noop {} {\bibfield  {journal} {\bibinfo  {journal}
  {Nat. Phys.}\ }\textbf {\bibinfo {volume} {6}},\ \bibinfo {pages} {480}
  (\bibinfo {year} {2010})}\BibitemShut {NoStop}%
\bibitem [{\citenamefont {Nicolis}\ and\ \citenamefont
  {Prigogine}(1977)}]{NicolisPrigogine}%
  \BibitemOpen
  \bibfield  {author} {\bibinfo {author} {\bibfnamefont {G.}~\bibnamefont
  {Nicolis}}\ and\ \bibinfo {author} {\bibfnamefont {I.}~\bibnamefont
  {Prigogine}},\ }\href@noop {} {\emph {\bibinfo {title} {Self-organization in
  nonequiibrium systems: From dissipative structures to order through
  fluctuations}}}\ (\bibinfo  {publisher} {J. Wiley and Sons},\ \bibinfo {year}
  {1977})\BibitemShut {NoStop}%
\bibitem [{\citenamefont {Murray}(2001)}]{Murray}%
  \BibitemOpen
  \bibfield  {author} {\bibinfo {author} {\bibfnamefont {J.}~\bibnamefont
  {Murray}},\ }\href@noop {} {\emph {\bibinfo {title} {Mathematical biology II:
  Spatial models and biomedical applications}}}\ (\bibinfo  {publisher}
  {Springer-Verlag},\ \bibinfo {year} {2001})\BibitemShut {NoStop}%
\bibitem [{\citenamefont {Nakao}\ and\ \citenamefont
  {Mikhailov}(2010)}]{NakaoMikhailov}%
  \BibitemOpen
  \bibfield  {author} {\bibinfo {author} {\bibfnamefont {H.}~\bibnamefont
  {Nakao}}\ and\ \bibinfo {author} {\bibfnamefont {A.}~\bibnamefont
  {Mikhailov}},\ }\href@noop {} {\bibfield  {journal} {\bibinfo  {journal}
  {Nat. Phys.}\ }\textbf {\bibinfo {volume} {6}},\ \bibinfo {pages} {544}
  (\bibinfo {year} {2010})}\BibitemShut {NoStop}%
\bibitem [{\citenamefont {Turing}(1952)}]{Turing}%
  \BibitemOpen
  \bibfield  {author} {\bibinfo {author} {\bibfnamefont {A.}~\bibnamefont
  {Turing}},\ }\href@noop {} {\bibfield  {journal} {\bibinfo  {journal} {Phil.
  Trans. R. Soc. Lond. B}\ }\textbf {\bibinfo {volume} {237}},\ \bibinfo
  {pages} {37} (\bibinfo {year} {1952})}\BibitemShut {NoStop}%
\bibitem [{\citenamefont {Castets}\ \emph {et~al.}(1990)\citenamefont
  {Castets}, \citenamefont {Dulos}, \citenamefont {Boissonade},\ and\
  \citenamefont {De~Kepper}}]{Castets}%
  \BibitemOpen
  \bibfield  {author} {\bibinfo {author} {\bibfnamefont {V.}~\bibnamefont
  {Castets}}, \bibinfo {author} {\bibfnamefont {E.}~\bibnamefont {Dulos}},
  \bibinfo {author} {\bibfnamefont {J.}~\bibnamefont {Boissonade}}, \ and\
  \bibinfo {author} {\bibfnamefont {P.}~\bibnamefont {De~Kepper}},\ }\href@noop
  {} {\bibfield  {journal} {\bibinfo  {journal} {Phys. Rev. Lett.}\ }\textbf
  {\bibinfo {volume} {64}},\ \bibinfo {pages} {2953} (\bibinfo {year}
  {1990})}\BibitemShut {NoStop}%
\bibitem [{\citenamefont {De~Kepper}\ \emph {et~al.}(1991)\citenamefont
  {De~Kepper}, \citenamefont {Castets}, \citenamefont {Dulos},\ and\
  \citenamefont {Boissonade}}]{DeKepper}%
  \BibitemOpen
  \bibfield  {author} {\bibinfo {author} {\bibfnamefont {P.}~\bibnamefont
  {De~Kepper}}, \bibinfo {author} {\bibfnamefont {V.}~\bibnamefont {Castets}},
  \bibinfo {author} {\bibfnamefont {E.}~\bibnamefont {Dulos}}, \ and\ \bibinfo
  {author} {\bibfnamefont {J.}~\bibnamefont {Boissonade}},\ }\href@noop {}
  {\bibfield  {journal} {\bibinfo  {journal} {Physica D}\ }\textbf {\bibinfo
  {volume} {49}} (\bibinfo {year} {1991})}\BibitemShut {NoStop}%
\bibitem [{\citenamefont {Tompkins}\ \emph {et~al.}(2014)\citenamefont
  {Tompkins}, \citenamefont {Li}, \citenamefont {Girabawe}, \citenamefont
  {Heymann}, \citenamefont {Ermentrout}, \citenamefont {Epstein},\ and\
  \citenamefont {Fraden}}]{Tompkins}%
  \BibitemOpen
  \bibfield  {author} {\bibinfo {author} {\bibfnamefont {N.}~\bibnamefont
  {Tompkins}}, \bibinfo {author} {\bibfnamefont {N.}~\bibnamefont {Li}},
  \bibinfo {author} {\bibfnamefont {C.}~\bibnamefont {Girabawe}}, \bibinfo
  {author} {\bibfnamefont {M.}~\bibnamefont {Heymann}}, \bibinfo {author}
  {\bibfnamefont {G.}~\bibnamefont {Ermentrout}}, \bibinfo {author}
  {\bibfnamefont {I.}~\bibnamefont {Epstein}}, \ and\ \bibinfo {author}
  {\bibfnamefont {S.}~\bibnamefont {Fraden}},\ }\href@noop {} {\bibfield
  {journal} {\bibinfo  {journal} {Proc. Natl. Acad. Sci. U.S.A.}\ }\textbf
  {\bibinfo {volume} {111}},\ \bibinfo {pages} {4397} (\bibinfo {year}
  {2014})}\BibitemShut {NoStop}%
\bibitem [{\citenamefont {Pismen}(2006)}]{Pismen}%
  \BibitemOpen
  \bibfield  {author} {\bibinfo {author} {\bibfnamefont {L.}~\bibnamefont
  {Pismen}},\ }\href@noop {} {\emph {\bibinfo {title} {Patterns and interfaces
  in dissipative dynamics}}}\ (\bibinfo  {publisher} {Springer Science \&
  Business Media},\ \bibinfo {year} {2006})\BibitemShut {NoStop}%
\bibitem [{\citenamefont {Gierer}\ and\ \citenamefont
  {Meinhardt}(1972)}]{GiererMeinhardt}%
  \BibitemOpen
  \bibfield  {author} {\bibinfo {author} {\bibfnamefont {A.}~\bibnamefont
  {Gierer}}\ and\ \bibinfo {author} {\bibfnamefont {H.}~\bibnamefont
  {Meinhardt}},\ }\href@noop {} {\bibfield  {journal} {\bibinfo  {journal}
  {Kybernetik}\ }\textbf {\bibinfo {volume} {12}},\ \bibinfo {pages} {30}
  (\bibinfo {year} {1972})}\BibitemShut {NoStop}%
\bibitem [{\citenamefont {Asllani}\ \emph {et~al.}(2014)\citenamefont
  {Asllani}, \citenamefont {Challenger}, \citenamefont {Pavone}, \citenamefont
  {Sacconi},\ and\ \citenamefont {Fanelli}}]{Asllani1}%
  \BibitemOpen
  \bibfield  {author} {\bibinfo {author} {\bibfnamefont {M.}~\bibnamefont
  {Asllani}}, \bibinfo {author} {\bibfnamefont {J.}~\bibnamefont {Challenger}},
  \bibinfo {author} {\bibfnamefont {F.}~\bibnamefont {Pavone}}, \bibinfo
  {author} {\bibfnamefont {L.}~\bibnamefont {Sacconi}}, \ and\ \bibinfo
  {author} {\bibfnamefont {D.}~\bibnamefont {Fanelli}},\ }\href@noop {}
  {\bibfield  {journal} {\bibinfo  {journal} {Nature Communication}\ }\textbf
  {\bibinfo {volume} {5}} (\bibinfo {year} {2014})}\BibitemShut {NoStop}%
\bibitem [{\citenamefont {Carletti}\ and\ \citenamefont
  {Muolo}(2022)}]{carletti}%
  \BibitemOpen
  \bibfield  {author} {\bibinfo {author} {\bibfnamefont {T.}~\bibnamefont
  {Carletti}}\ and\ \bibinfo {author} {\bibfnamefont {R.}~\bibnamefont
  {Muolo}},\ }\href@noop {} {\bibfield  {journal} {\bibinfo  {journal} {Chaos
  Solit. Fractals}\ }\textbf {\bibinfo {volume} {164}},\ \bibinfo {pages}
  {112638} (\bibinfo {year} {2022})}\BibitemShut {NoStop}%
\bibitem [{\citenamefont {Busiello}\ \emph {et~al.}(2015)\citenamefont
  {Busiello}, \citenamefont {Planchon}, \citenamefont {Asllani}, \citenamefont
  {Carletti},\ and\ \citenamefont {Fanelli}}]{busiello_turing}%
  \BibitemOpen
  \bibfield  {author} {\bibinfo {author} {\bibfnamefont {D.}~\bibnamefont
  {Busiello}}, \bibinfo {author} {\bibfnamefont {G.}~\bibnamefont {Planchon}},
  \bibinfo {author} {\bibfnamefont {M.}~\bibnamefont {Asllani}}, \bibinfo
  {author} {\bibfnamefont {T.}~\bibnamefont {Carletti}}, \ and\ \bibinfo
  {author} {\bibfnamefont {D.}~\bibnamefont {Fanelli}},\ }\href@noop {}
  {\bibfield  {journal} {\bibinfo  {journal} {Eur. Phys. J. B}\ }\textbf
  {\bibinfo {volume} {88}},\ \bibinfo {pages} {222} (\bibinfo {year}
  {2015})}\BibitemShut {NoStop}%
\bibitem [{\citenamefont {Asllani}\ \emph {et~al.}(2016)\citenamefont
  {Asllani}, \citenamefont {Carletti},\ and\ \citenamefont
  {Fanelli}}]{Asllani2016}%
  \BibitemOpen
  \bibfield  {author} {\bibinfo {author} {\bibfnamefont {M.}~\bibnamefont
  {Asllani}}, \bibinfo {author} {\bibfnamefont {T.}~\bibnamefont {Carletti}}, \
  and\ \bibinfo {author} {\bibfnamefont {D.}~\bibnamefont {Fanelli}},\
  }\href@noop {} {\bibfield  {journal} {\bibinfo  {journal} {Eur. Phys. J. B}\
  ,\ \bibinfo {pages} {89}} (\bibinfo {year} {2016})}\BibitemShut {NoStop}%
\bibitem [{\citenamefont {Petit}\ \emph {et~al.}(2017)\citenamefont {Petit},
  \citenamefont {Lauwens}, \citenamefont {Fanelli},\ and\ \citenamefont
  {Carletti}}]{PABFC2017}%
  \BibitemOpen
  \bibfield  {author} {\bibinfo {author} {\bibfnamefont {J.}~\bibnamefont
  {Petit}}, \bibinfo {author} {\bibfnamefont {B.}~\bibnamefont {Lauwens}},
  \bibinfo {author} {\bibfnamefont {D.}~\bibnamefont {Fanelli}}, \ and\
  \bibinfo {author} {\bibfnamefont {T.}~\bibnamefont {Carletti}},\ }\href@noop
  {} {\bibfield  {journal} {\bibinfo  {journal} {Phys. Rev. Lett.}\ }\textbf
  {\bibinfo {volume} {119}},\ \bibinfo {pages} {148301} (\bibinfo {year}
  {2017})}\BibitemShut {NoStop}%
\bibitem [{\citenamefont {Muolo}\ \emph {et~al.}(2023)\citenamefont {Muolo},
  \citenamefont {Gallo}, \citenamefont {Latora}, \citenamefont {Frasca},\ and\
  \citenamefont {Carletti}}]{muologallo}%
  \BibitemOpen
  \bibfield  {author} {\bibinfo {author} {\bibfnamefont {R.}~\bibnamefont
  {Muolo}}, \bibinfo {author} {\bibfnamefont {L.}~\bibnamefont {Gallo}},
  \bibinfo {author} {\bibfnamefont {V.}~\bibnamefont {Latora}}, \bibinfo
  {author} {\bibfnamefont {M.}~\bibnamefont {Frasca}}, \ and\ \bibinfo {author}
  {\bibfnamefont {T.}~\bibnamefont {Carletti}},\ }\href@noop {} {\bibfield
  {journal} {\bibinfo  {journal} {Chaos Solit. Fractals}\ }\textbf {\bibinfo
  {volume} {166}},\ \bibinfo {pages} {112912} (\bibinfo {year}
  {2023})}\BibitemShut {NoStop}%
\bibitem [{\citenamefont {Giambagli}\ \emph {et~al.}(2022)\citenamefont
  {Giambagli}, \citenamefont {Calmon}, \citenamefont {Muolo}, \citenamefont
  {Carletti},\ and\ \citenamefont {Bianconi}}]{turing_topological}%
  \BibitemOpen
  \bibfield  {author} {\bibinfo {author} {\bibfnamefont {L.}~\bibnamefont
  {Giambagli}}, \bibinfo {author} {\bibfnamefont {M.}~\bibnamefont {Calmon}},
  \bibinfo {author} {\bibfnamefont {R.}~\bibnamefont {Muolo}}, \bibinfo
  {author} {\bibfnamefont {T.}~\bibnamefont {Carletti}}, \ and\ \bibinfo
  {author} {\bibfnamefont {G.}~\bibnamefont {Bianconi}},\ }\href@noop {}
  {\bibfield  {journal} {\bibinfo  {journal} {Phys. Rev. E}\ }\textbf {\bibinfo
  {volume} {106}} (\bibinfo {year} {2022})}\BibitemShut {NoStop}%
\bibitem [{\citenamefont {Asllani}\ \emph {et~al.}(2018)\citenamefont
  {Asllani}, \citenamefont {Lambiotte},\ and\ \citenamefont
  {Carletti}}]{malbor_teo}%
  \BibitemOpen
  \bibfield  {author} {\bibinfo {author} {\bibfnamefont {M.}~\bibnamefont
  {Asllani}}, \bibinfo {author} {\bibfnamefont {R.}~\bibnamefont {Lambiotte}},
  \ and\ \bibinfo {author} {\bibfnamefont {T.}~\bibnamefont {Carletti}},\
  }\href@noop {} {\bibfield  {journal} {\bibinfo  {journal} {Sci. Adv.}\
  }\textbf {\bibinfo {volume} {4}},\ \bibinfo {pages} {Eaau9403} (\bibinfo
  {year} {2018})}\BibitemShut {NoStop}%
\bibitem [{\citenamefont {Duan}\ \emph {et~al.}(2022)\citenamefont {Duan},
  \citenamefont {Nishikawa}, \citenamefont {Eroglu},\ and\ \citenamefont
  {Motter}}]{DNEM}%
  \BibitemOpen
  \bibfield  {author} {\bibinfo {author} {\bibfnamefont {C.}~\bibnamefont
  {Duan}}, \bibinfo {author} {\bibfnamefont {T.}~\bibnamefont {Nishikawa}},
  \bibinfo {author} {\bibfnamefont {D.}~\bibnamefont {Eroglu}}, \ and\ \bibinfo
  {author} {\bibfnamefont {A.~E.}\ \bibnamefont {Motter}},\ }\href@noop {}
  {\bibfield  {journal} {\bibinfo  {journal} {Science Advances}\ }\textbf
  {\bibinfo {volume} {8}},\ \bibinfo {pages} {eabm8310} (\bibinfo {year}
  {2022})}\BibitemShut {NoStop}%
\bibitem [{\citenamefont {Trefethen}\ and\ \citenamefont
  {Embree}(2005)}]{trefethen}%
  \BibitemOpen
  \bibfield  {author} {\bibinfo {author} {\bibfnamefont {L.}~\bibnamefont
  {Trefethen}}\ and\ \bibinfo {author} {\bibfnamefont {M.}~\bibnamefont
  {Embree}},\ }\href@noop {} {\emph {\bibinfo {title} {Spectra and
  Pseudospectra: The Behavior of Nonnormal Matrices and Operators}}}\ (\bibinfo
   {publisher} {Princeton Univ. Press},\ \bibinfo {year} {2005})\BibitemShut
  {NoStop}%
\bibitem [{\citenamefont {Asllani}\ and\ \citenamefont
  {Carletti}(2018)}]{top_resilience}%
  \BibitemOpen
  \bibfield  {author} {\bibinfo {author} {\bibfnamefont {M.}~\bibnamefont
  {Asllani}}\ and\ \bibinfo {author} {\bibfnamefont {T.}~\bibnamefont
  {Carletti}},\ }\href@noop {} {\bibfield  {journal} {\bibinfo  {journal}
  {Phys. Rev. E}\ }\textbf {\bibinfo {volume} {97}} (\bibinfo {year}
  {2018})}\BibitemShut {NoStop}%
\bibitem [{\citenamefont {Muolo}\ \emph {et~al.}(2019)\citenamefont {Muolo},
  \citenamefont {Asllani}, \citenamefont {Fanelli}, \citenamefont {Maini},\
  and\ \citenamefont {Carletti}}]{jtb}%
  \BibitemOpen
  \bibfield  {author} {\bibinfo {author} {\bibfnamefont {R.}~\bibnamefont
  {Muolo}}, \bibinfo {author} {\bibfnamefont {M.}~\bibnamefont {Asllani}},
  \bibinfo {author} {\bibfnamefont {D.}~\bibnamefont {Fanelli}}, \bibinfo
  {author} {\bibfnamefont {P.}~\bibnamefont {Maini}}, \ and\ \bibinfo {author}
  {\bibfnamefont {T.}~\bibnamefont {Carletti}},\ }\href@noop {} {\bibfield
  {journal} {\bibinfo  {journal} {J. Theor. Biol.}\ }\textbf {\bibinfo {volume}
  {480}},\ \bibinfo {pages} {81} (\bibinfo {year} {2019})}\BibitemShut
  {NoStop}%
\bibitem [{\citenamefont {Nishikawa}\ and\ \citenamefont
  {Motter}(2006)}]{mott_nish2}%
  \BibitemOpen
  \bibfield  {author} {\bibinfo {author} {\bibfnamefont {T.}~\bibnamefont
  {Nishikawa}}\ and\ \bibinfo {author} {\bibfnamefont {A.}~\bibnamefont
  {Motter}},\ }\href@noop {} {\bibfield  {journal} {\bibinfo  {journal} {Phys.
  Rev. E}\ }\textbf {\bibinfo {volume} {73}},\ \bibinfo {pages} {065106(R)}
  (\bibinfo {year} {2006})}\BibitemShut {NoStop}%
\bibitem [{\citenamefont {Muolo}\ \emph {et~al.}(2021)\citenamefont {Muolo},
  \citenamefont {Carletti}, \citenamefont {Gleeson},\ and\ \citenamefont
  {Asllani}}]{entropy}%
  \BibitemOpen
  \bibfield  {author} {\bibinfo {author} {\bibfnamefont {R.}~\bibnamefont
  {Muolo}}, \bibinfo {author} {\bibfnamefont {T.}~\bibnamefont {Carletti}},
  \bibinfo {author} {\bibfnamefont {J.}~\bibnamefont {Gleeson}}, \ and\
  \bibinfo {author} {\bibfnamefont {M.}~\bibnamefont {Asllani}},\ }\href@noop
  {} {\bibfield  {journal} {\bibinfo  {journal} {Entropy}\ }\textbf {\bibinfo
  {volume} {23}},\ \bibinfo {pages} {36} (\bibinfo {year} {2021})}\BibitemShut
  {NoStop}%
\bibitem [{\citenamefont {Prigogine}\ and\ \citenamefont
  {Nicolis}(1967)}]{PrigogineNicolis1967}%
  \BibitemOpen
  \bibfield  {author} {\bibinfo {author} {\bibfnamefont {I.}~\bibnamefont
  {Prigogine}}\ and\ \bibinfo {author} {\bibfnamefont {G.}~\bibnamefont
  {Nicolis}},\ }\href@noop {} {\bibfield  {journal} {\bibinfo  {journal} {J.
  Chem. Phys.}\ }\textbf {\bibinfo {volume} {46}},\ \bibinfo {pages} {3542}
  (\bibinfo {year} {1967})}\BibitemShut {NoStop}%
\bibitem [{\citenamefont {Prigogine}\ and\ \citenamefont
  {Lefever}(1968)}]{PrigogineLefever1968}%
  \BibitemOpen
  \bibfield  {author} {\bibinfo {author} {\bibfnamefont {I.}~\bibnamefont
  {Prigogine}}\ and\ \bibinfo {author} {\bibfnamefont {R.}~\bibnamefont
  {Lefever}},\ }\href@noop {} {\bibfield  {journal} {\bibinfo  {journal} {J.
  Chem. Phys.}\ }\textbf {\bibinfo {volume} {48}},\ \bibinfo {pages} {1695}
  (\bibinfo {year} {1968})}\BibitemShut {NoStop}%
\bibitem [{\citenamefont {Boland}\ \emph {et~al.}(2008)\citenamefont {Boland},
  \citenamefont {Galla},\ and\ \citenamefont {McKane}}]{galla}%
  \BibitemOpen
  \bibfield  {author} {\bibinfo {author} {\bibfnamefont {R.}~\bibnamefont
  {Boland}}, \bibinfo {author} {\bibfnamefont {T.}~\bibnamefont {Galla}}, \
  and\ \bibinfo {author} {\bibfnamefont {A.}~\bibnamefont {McKane}},\
  }\href@noop {} {\bibfield  {journal} {\bibinfo  {journal} {J. Stat. Mech.}\
  ,\ \bibinfo {pages} {P09001}} (\bibinfo {year} {2008})}\BibitemShut {NoStop}%
\bibitem [{\citenamefont {Koseska}\ \emph {et~al.}(2013)\citenamefont
  {Koseska}, \citenamefont {Volkov},\ and\ \citenamefont {Kurths}}]{KVK2013}%
  \BibitemOpen
  \bibfield  {author} {\bibinfo {author} {\bibfnamefont {A.}~\bibnamefont
  {Koseska}}, \bibinfo {author} {\bibfnamefont {E.}~\bibnamefont {Volkov}}, \
  and\ \bibinfo {author} {\bibfnamefont {J.}~\bibnamefont {Kurths}},\
  }\href@noop {} {\bibfield  {journal} {\bibinfo  {journal} {Phys. Rep.}\
  }\textbf {\bibinfo {volume} {531}},\ \bibinfo {pages} {173} (\bibinfo {year}
  {2013})}\BibitemShut {NoStop}%
\bibitem [{\citenamefont {Lucas}\ \emph {et~al.}(2018)\citenamefont {Lucas},
  \citenamefont {Fanelli}, \citenamefont {Carletti},\ and\ \citenamefont
  {Petit}}]{LFCP2018}%
  \BibitemOpen
  \bibfield  {author} {\bibinfo {author} {\bibfnamefont {M.}~\bibnamefont
  {Lucas}}, \bibinfo {author} {\bibfnamefont {D.}~\bibnamefont {Fanelli}},
  \bibinfo {author} {\bibfnamefont {T.}~\bibnamefont {Carletti}}, \ and\
  \bibinfo {author} {\bibfnamefont {J.}~\bibnamefont {Petit}},\ }\href@noop {}
  {\bibfield  {journal} {\bibinfo  {journal} {EPL}\ }\textbf {\bibinfo {volume}
  {121}},\ \bibinfo {pages} {58008} (\bibinfo {year} {2018})}\BibitemShut
  {NoStop}%
\bibitem [{\citenamefont {O'Brien}\ \emph {et~al.}(2021)\citenamefont
  {O'Brien}, \citenamefont {Oliveira}, \citenamefont {Gleeson},\ and\
  \citenamefont {Asllani}}]{asllani_leaders}%
  \BibitemOpen
  \bibfield  {author} {\bibinfo {author} {\bibfnamefont {J.}~\bibnamefont
  {O'Brien}}, \bibinfo {author} {\bibfnamefont {K.}~\bibnamefont {Oliveira}},
  \bibinfo {author} {\bibfnamefont {J.}~\bibnamefont {Gleeson}}, \ and\
  \bibinfo {author} {\bibfnamefont {M.}~\bibnamefont {Asllani}},\ }\href@noop
  {} {\bibfield  {journal} {\bibinfo  {journal} {Phys. Rev. Res.}\ }\textbf
  {\bibinfo {volume} {3}},\ \bibinfo {pages} {023117} (\bibinfo {year}
  {2021})}\BibitemShut {NoStop}%
\bibitem [{\citenamefont {Forrow}\ \emph {et~al.}(2018)\citenamefont {Forrow},
  \citenamefont {Woodhouse},\ and\ \citenamefont {Dunkel}}]{FWD2018}%
  \BibitemOpen
  \bibfield  {author} {\bibinfo {author} {\bibfnamefont {A.}~\bibnamefont
  {Forrow}}, \bibinfo {author} {\bibfnamefont {F.}~\bibnamefont {Woodhouse}}, \
  and\ \bibinfo {author} {\bibfnamefont {J.}~\bibnamefont {Dunkel}},\
  }\href@noop {} {\bibfield  {journal} {\bibinfo  {journal} {Phys. Rev. X}\
  }\textbf {\bibinfo {volume} {8}},\ \bibinfo {pages} {041043} (\bibinfo {year}
  {2018})}\BibitemShut {NoStop}%
\bibitem [{\citenamefont {Nicoletti}\ \emph {et~al.}(2021)\citenamefont
  {Nicoletti}, \citenamefont {Carletti}, \citenamefont {Fanelli}, \citenamefont
  {Battistelli},\ and\ \citenamefont {Chisci}}]{NCFBC}%
  \BibitemOpen
  \bibfield  {author} {\bibinfo {author} {\bibfnamefont {S.}~\bibnamefont
  {Nicoletti}}, \bibinfo {author} {\bibfnamefont {T.}~\bibnamefont {Carletti}},
  \bibinfo {author} {\bibfnamefont {D.}~\bibnamefont {Fanelli}}, \bibinfo
  {author} {\bibfnamefont {G.}~\bibnamefont {Battistelli}}, \ and\ \bibinfo
  {author} {\bibfnamefont {L.}~\bibnamefont {Chisci}},\ }\href@noop {}
  {\bibfield  {journal} {\bibinfo  {journal} {J. Phys. Complex.}\ }\textbf
  {\bibinfo {volume} {2}} (\bibinfo {year} {2021})}\BibitemShut {NoStop}%
\end{thebibliography}%

\onecolumngrid
\appendix
\section{Condition for the Turing instability.}
\label{sec:instreg}

In the main text we showed (see Eqs.~\eqref{eq:lincompactJordanj1}-\eqref{eq:lincompactJordanjmj}) that Turing instability can be determined by studying the spectrum of the matrix $\mathbf{M}_j=\mathbf{J}_0+\Lambda^{(j)} \mathbf{D}$, where $\mathbf{J}_0$ is the Jacobian of the reaction part~\eqref{eq:J0} evaluated at the homogeneous equilibrium $(u_*,v_*)$, $\mathbf{D}$ is the diagonal matrix of the diffusion coefficients and $\Lambda^{(j)}$ is the $j$-th eigenvalue of the Laplace matrix $\mathbf{L}$. Moreover because the homogeneous solution is stable we also have $\mathrm{tr}(\mathbf{J}_0)<0$ and $\det \mathbf{J}_0>0$.

The eigenvalues of $\mathbf{M}_j$ can be straightforwardly obtained by using the formula
\begin{equation*}
 \mu_j=\frac{\mathrm{tr}(\mathbf{M}_j)+\sqrt{(\mathrm{tr}(\mathbf{M}_j))^2-4\det \mathbf{M}_j}}{2}\, ;
\end{equation*}
because the network is directed the Laplace matrix is generally asymmetric, thus its eigenvalues are complex numbers and so does $\mu_j$. By writing $\Lambda^{(j)}=\Re\Lambda^{(j)}+i\Im \Lambda^{(j)}$ we can obtain
\begin{eqnarray*}
 \Re\mathrm{tr}(\mathbf{M}_j)&=& \mathrm{tr}(\mathbf{J}_0)+(D_u+D_v)\Re\Lambda^{(j)}\\
 \Im\mathrm{tr}(\mathbf{M}_j)&=& (D_u+D_v)\Im\Lambda^{(j)}\\ 
 \Re\det(\mathbf{M}_j)&=&\det(\mathbf{J}_0)+\left(({J}_0)_{11}D_v+({J}_0)_{22}D_u\right)\Re\Lambda^{(j)}+D_uD_v\left[(\Re\Lambda^{(j)})^2-(\Im\Lambda^{(j)})^2\right]\\
 \Im\det(\mathbf{M}_j)&=& \left(({J}_0)_{11}D_v+({J}_0)_{22}D_u\right)\Im\Lambda^{(j)}+2D_uD_v\Re\Lambda^{(j)}\Im\Lambda^{(j)}\, ,
\end{eqnarray*}
and eventually express the real part of $\mu_j$ as follows
\begin{equation*}
 \Re\mu_j=\frac{\Re\mathrm{tr}(\mathbf{M}_j) + \gamma}{2}\, ,
\end{equation*}
where
\begin{equation*}
 \gamma=\sqrt{\frac{A+\sqrt{A^2+B^2}}{2}}\, , A=\left(\Re\mathrm{tr}(\mathbf{M}_j)\right)^2-\left(\Im\mathrm{tr}(\mathbf{M}_j)\right)^2-4\Re\det(\mathbf{M}_j)\text{ and } B=2 \Re\mathrm{tr}(\mathbf{M}_j) \Im\mathrm{tr}(\mathbf{M}_j)-4\Re\det(\mathbf{M}_j)\, .
\end{equation*}

The condition for instability is $\Re\mu_j>0$ for some $j$, namely
\begin{equation*}
\Re\mathrm{tr}(\mathbf{M}_j) > -\gamma\, ,
\end{equation*}
that can be rewritten as
\begin{equation}
S_2\left( \Re\Lambda^{(j)}\right)\left(\Im\Lambda^{(j)}\right)^2< -S_1\left(\Re\Lambda^{(j)}\right)\, ,
\label{eq:S2S1}
\end{equation}
where $S_1$ (resp. $S_2$) is a polynomial of fourth (resp. second) degree in $\Re\Lambda^{(j)}$. More precisely
\begin{eqnarray*}
S_1\left( x\right) &=&C_{14}x^4+C_{13}x^3+C_{12}x^2+C_{11}x+C_{10}\\
S_2\left( x\right) &=&C_{22}x^2+C_{21}x+C_{20}\, ,
\end{eqnarray*}
where the coefficients $C_{ij}$ are explicitly given by
\begin{eqnarray*}
C_{14} &=&D_uD_v(D_u+D_v)^2\\
C_{13} &=&(D_u+D_v)^2\left( (J_0)_{11}D_v+(J_0)_{22}D_u\right)+2\mathrm{tr}(\mathbf{J}_0)D_uD_v(D_u+D_v)\\
C_{12} &=&\det(\mathbf{J}_0)(D_u+D_v)^2+\left(\mathrm{tr}(\mathbf{J}_0)\right)^2D_uD_v+2\mathrm{tr}(\mathbf{J}_0)(D_u+D_v)\left( (J_0)_{11}D_v+(J_0)_{22}D_u\right)\\
C_{11} &=&2\mathrm{tr}(\mathbf{J}_0)(D_u+D_v)\det(\mathbf{J}_0)+\left(\mathrm{tr}(\mathbf{J}_0)\right)^2\left( (J_0)_{11}D_v+(J_0)_{22}D_u\right)\\
C_{10}&=&\det(\mathbf{J}_0)\left(\mathrm{tr}(\mathbf{J}_0)\right)^2\, ,
\end{eqnarray*}
and
\begin{eqnarray*}
C_{22} &=&D_uD_v(D_u-D_v)^2\\
C_{21} &=&\left( (J_0)_{11}D_v+(J_0)_{22}D_u\right)(D_u-D_v)^2\\
C_{20}&=&(J_0)_{11}(J_0)_{22}(D_u-D_v)^2\, .
\end{eqnarray*}

In conclusion, whenever the Laplace matrix admits at least one eigenvalue such that condition~\eqref{eq:S2S1} is satisfied, then the homogeneous equilibrium turns out to be unstable once submitted to heterogeneous perturbations. \textcolor{black}{The region of the complex plane $(\Re \Lambda,\Im\Lambda)$ where this condition holds true is the green region shown in Fig.~\ref{fig:Reginst} in the main text and Fig.~\ref{fig:ReginstApp} in Appendix~\ref{eq:otherex}.}

\section{Pattern reconstruction}
\label{sec:recontr}
To  reconstruct the pattern, $u(\hat{t})$, we first centered it by subtracting the homogeneous equilibrium $u_*$. The goal is thus to project the obtained vector, ${p}=u(\hat{t})-u_*\in \mathbb{R}^n$, on the subspace $V$ generated by the $d\geq 1$ linearly independent vectors considered for the reconstruction, i.e., unstable eigenvectors and generalized ones. Let us denote by $\pi_{V}({p})$ the projection of ${p}$ onto $V$. This vector can thus be expressed by a linear combination of those generating vectors
\begin{equation}
\label{eq:projVp}
    \pi_{V}({p})=\mathbf{B}{a}\, ,
\end{equation}
where ${a}$ is the vector containing the coefficients of the linear combination and $\mathbf{B}$ is the matrix with the $d$ linearly independent vectors as columns. By using basic algebra, one can express the centered pattern as ${p}=\pi_{V}({p})+{z}$, with $\pi_{V}({p}) \in V$ and ${z} \in V^{\perp}$. By invoking the image-kernel theorem one can write $V^\perp=\Im(\mathbf{B})^\perp= \ker(\mathbf{B}^\top)$, $z={p} - \pi_{V}({p})$ lies therefore in $\ker(\mathbf{B}^\top)$, namely $\mathbf{B}^\top({p} - \pi_{V}({p}))= {0}$. By using the expression~\eqref{eq:projVp} we can rewrite the latter as
\begin{equation*}
\mathbf{B}^\top({p} - \mathbf{B} {a})= {0}\, .
\end{equation*}
By developing the computations we obtain
\begin{equation*}
{a}=(\mathbf{B}^\top \mathbf{B})^{-1} \mathbf{B}^\top {p}\, ,
\end{equation*}
and eventually the following expression for the projected centered pattern $\pi_{V}({p})= \mathbf{B} (\mathbf{B}^\top \mathbf{B})^{-1} \mathbf{B}^\top {p}$. Let us observe that if the (generalized) eigenvectors are complex, and so does the matrix $\mathbf{B}$, we decided to replace each complex vector with two real ones obtained by taking the real and the imaginary part of the former complex vector.

For sake of clarity in the Figures presenting the pattern, the equilibrium has been added back to the projection to better compare $\pi_{V}({p})+u_*$ with the pattern $u(\hat{t})$.

We can now focus more on the vectors we use for the reconstruction. As said in the main text, when the Laplace matrix is diagonalizable we just compute the eigenvectors and use those associated with unstable eigenvalues. When we work with defective networks, the Laplace matrix does not have a linearly independent set of eigenvectors. We therefore compute the Jordan canonical form of the Laplace matrix, as seen in the main text, $\mathbf{P}^{-1}\mathbf{L}\mathbf{P}=\mathbf{B}=\mathrm{diag}(\mathbf{B}_1,\dots,\mathbf{B}_\ell)\,$  where the $\mathbf{B}_j$ is the $m_j\times m_j$ Jordan block, $m_1+\dots+m_\ell=n$. The matrix $\mathbf{P}$ has the (generalized) eigenvectors as columns, which are linearly independent. We then use the (generalized) eigenvectors corresponding to unstable eigenvalues in the reconstruction.

In order to compare the reconstructions with and without the generalized eigenvectors, we used a weighted absolute error. If we note by $\tilde{u}=(\tilde{u}_1,\dots,\tilde{u}_{n})^\top=\pi_{V}({p})+u_*$ the reconstructed pattern starting from the real one $\hat{u}=(u_1(\hat{t}),\dots,u_{n}(\hat{t}))^\top$ for the fixed time $\hat{t}$, the absolute error is $\| \hat{u} - \tilde{u} \|_1/n$, i.e., the $1$-norm of the $n$-dimensional vectors divided by the network size. The latter error is then weighted by the number of (generalized) eigenvectors used in the reconstruction, i.e., the dimension $d$ of the subspace $V$, divided by the total number of (generalized) unstable available eigenvectors, $N_u$. In conclusion the error used to evaluate the goodness of the reconstructed pattern is
\begin{equation}
\label{eq:werrordef}
 \varepsilon = \frac{d}{N_u}\frac{\| \hat{u} - \tilde{u} \|_1}{n} \, .
\end{equation}

{\color{black}{
\section{Generating directed networks with prescribed defective Laplacian spectra}
\label{eq:dirdefnet}

The goal of this section is to present a novel algorithm allowing to determine a directed network with a prescribed defective Laplacian spectrum. To the best of our knowledge this is an open problem and in the literature few results are available, the interested reader can consult~\cite{FWD2018} for the case of symmetric networks or~\cite{NCFBC} in the case of directed ones, and the references therein. Let us observe that none of the previous works can be applied to the present case because both assume the existence of a basis for the Laplace matrix. For the scope of this work we thus decided to develop the algorithm under the simplifying assumption of a real non-positive spectrum. 

Let us thus consider a collection of $s+1\geq 2$ real eigenvalues, $\Lambda^{(0)}=0$, $\Lambda^{(j)}<0$ and let us assume moreover that each eigenvalue has algebraic multiplicity $m_j \geq 1$, for $j=1,\dots,s$, strictly larger than the geometric multiplicity. Let us initially impose the null eigenvalue to be simple, i.e., $m_0=1$; we will relax this assumption in the following. We thus trivially have $\sum_{j=1}^s m_j = n$ and the network will thus have $n+1$ nodes.

To each eigenvalue $\Lambda^{(j)}$, $j=1,\dots, s$, we associate a Jordan block of dimension $m_j\times m_j$
\begin{equation}
\label{eq:Bj}
 \mathbf{B}^{(j)}=\Lambda^{(j)}\mathbf{I}_{m_j} + \mathbf{N}_{m_j}\, ,
\end{equation}
where $\mathbf{I}_{m_j}$ is the $m_j\times m_j$ identity matrix and $\mathbf{N}_{m_j}$ the nilpotent matrix 
\begin{equation*}
 \mathbf{N}_{m_j}=\left(
\begin{matrix}
 0 & 1 & 0 & \dots &0\\
 0 & 0 & 1 & \dots&0\\
 \vdots & \vdots & \ddots & \ddots & \vdots\\
 0 & 0 & 0 & \ddots &1\\
 0 & 0 & 0 & 0 &0
\end{matrix}\right)\, .
\end{equation*}

Let us introduce the $(n+1)\times (n+1)$ block diagonal matrix $\mathbf{J}_L$
\begin{equation}
\label{eq:JLmat}
\mathbf{J}_L = \left(\begin{matrix}0 & \vec{0}_n^\top\\
\vec{0}_n & \mathbf{B}
\end{matrix}\right)\, ,
\end{equation}
where $\vec{0}_n=(0,\dots,0)^\top$ is the $n$-dimensional null vector and $\mathbf{B}$ the $n\times n$ block diagonal matrix build with the Jordan block, namely
\begin{equation}
\label{eq:B}
\mathbf{B}=\mathrm{diag}(\mathbf{B}^{(1)},\dots, \mathbf{B}^{(s)})\, .
\end{equation}
Let us observe that we do not consider in $\mathbf{B}$ the null eigenvalue that has been already set in the entry $(1,1)$ of $\mathbf{J}_L$. The latter matrix will be the Jordan Canonical Form of the Laplace matrix we are looking for.

Let us define the $(n+1)\times (n+1)$ non-singular matrix $\mathbf{S}$ given by
\begin{equation}
\label{eq:S}
\mathbf{S}=\left(
\begin{matrix}
 1 & \vec{0}_n^\top  \\
 \mathbf{u}_n & \mathbf{I}_n
\end{matrix}\right)\, ,
\end{equation}
where we have introduced the $n$-dimensional vector $ \mathbf{u}_n=(1,\dots,1)^\top$. One can easily prove that the inverse of $\mathbf{S}$ exists and it is given by
\begin{equation}
\label{eq:Sinv}
\mathbf{S}^{-1}=\left(
\begin{matrix}
 1 & \vec{0}_n^\top  \\
- \mathbf{u}_n & \mathbf{I}_n
\end{matrix}\right)\, .
\end{equation}

By using $\mathbf{S}$ and $\mathbf{S}^{-1}$, we define the matrix
\begin{equation}
\label{eq:L}
\mathbf{L}=\mathbf{S}\mathbf{J}_L\mathbf{S}^{-1}\, .
\end{equation}
Clearly this matrix has the same spectrum and same eigenvalues multiplicity of $\mathbf{J}_L$. It remains to prove that $\mathbf{L}$ is indeed a Laplace matrix of a suitable network whose adjacency matrix $\mathbf{A}$ is given by $A_{ij}=L_{ij}$ for $i\neq j$ and $A_{ii}=0$. In this way we will have built a network with a prescribed spectrum and eigenvalues multiplicities, and positively weighted adjacency matrix.

Let us introduce the $(n+1)$-dimensional vector $\mathbf{e}_1=(1,0,\dots,0)^\top$, then we trivially get
\begin{equation*}
 \mathbf{S}\mathbf{e}_1=\mathbf{u}_{n+1}\, ,
\end{equation*}
and thus
\begin{equation*}
 \mathbf{S}^{-1}\mathbf{u}_{n+1}=\mathbf{e}_1\, ,
\end{equation*}
hence
\begin{equation}
\label{eq:Lsum0}
\mathbf{L}\mathbf{u}_{n+1}=\mathbf{S}\mathbf{J}_L\mathbf{S}^{-1}\mathbf{u}_{n+1}=\mathbf{S}\mathbf{J}_L\mathbf{e}_1=0\, ,
\end{equation}
where in the last step we have used the block structure of the matrix $\mathbf{J}_L$ given by~\eqref{eq:JLmat}. We have thus proved that the rows of $\mathbf{L}$ sum to zero or equivalently that the constant vector $\mathbf{u}_{n+1}$ is an eigenvector associated to the eigenvalue $\Lambda^{(0)}=0$.

It remains to prove that $\mathbf{L}$ has non-positive diagonal and non-negative out of diagonal entries. To do this, let us compute explicitly $\mathbf{S}\mathbf{J}_L\mathbf{S}^{-1}$, moreover let us rewrite $\mathbf{S}$ and $\mathbf{S}^{-1}$ by using the block structure induced by the one of $\mathbf{B}$, namely
\begin{equation*}
\mathbf{L}=\mathbf{S}\mathbf{J}_L\mathbf{S}^{-1}=\left(
\begin{matrix}
 1 & \vec{0}_{m_1}^\top & \vec{0}_{m_2}^\top & \dots & \vec{0}_{m_s}^\top  \\
 \mathbf{u}_{m_1} & \mathbf{I}_{m_1} & \mathbf{O}_{m_1,m_2} & \dots & \mathbf{O}_{m_1,m_s} \\
 \mathbf{u}_{m_2} & \mathbf{O}_{m_2,m_1} & \mathbf{I}_{m_2} & \dots & \mathbf{O}_{m_2,m_s} \\
  \vdots & \vdots & \ddots & \ddots & \vdots \\
  \mathbf{u}_{m_s} & \mathbf{O}_{m_s,m_1} & \dots & \mathbf{O}_{m_s,m_{s-1}} & \mathbf{I}_{m_s}
 \end{matrix}\right)\left(\begin{matrix}0 & \vec{0}_N^\top\\
\vec{0}_N & \mathbf{B}
\end{matrix}\right)
\left(\begin{matrix}
 1 & \vec{0}_{m_1}^\top & \vec{0}_{m_2}^\top & \dots & \vec{0}_{m_s}^\top  \\
 -\mathbf{u}_{m_1} & \mathbf{I}_{m_1} & \mathbf{O}_{m_1,m_2} & \dots & \mathbf{O}_{m_1,m_s} \\
 -\mathbf{u}_{m_2} & \mathbf{O}_{m_2,m_1} & \mathbf{I}_{m_2} & \dots & \mathbf{O}_{m_2,m_s} \\
  \vdots & \vdots & \ddots & \ddots & \vdots \\
  -\mathbf{u}_{m_s} & \mathbf{O}_{m_s,m_1} & \dots & \mathbf{O}_{m_s,m_{s-1}} & \mathbf{I}_{m_s}
 \end{matrix}\right)\, ,
\end{equation*}
where we have introduced the null $m\times m'$ matrix denoted by $\mathbf{O}_{m,m'}$.

By performing the computation, we obtain
\begin{eqnarray*}
\mathbf{L}&=&\left(
\begin{matrix}
 0 & \vec{0}_{m_1}^\top & \vec{0}_{m_2}^\top & \dots & \vec{0}_{m_s}^\top  \\
\vec{0}_{m_1} & \mathbf{B}_{m_1} & \mathbf{O}_{m_1,m_2} & \dots & \mathbf{O}_{m_1,m_s} \\
\vec{0}_{m_2} & \mathbf{O}_{m_2,m_1} & \mathbf{B}_{m_2} & \dots & \mathbf{O}_{m_2,m_s} \\
  \vdots & \vdots & \ddots & \ddots & \vdots \\
\vec{0}_{m_s} & \mathbf{O}_{m_s,m_1} & \dots & \mathbf{O}_{m_s,m_{s-1}} & \mathbf{B}_{m_s}
 \end{matrix}\right)
 \left(\begin{matrix}
 1 & \vec{0}_{m_1}^\top & \vec{0}_{m_2}^\top & \dots & \vec{0}_{m_s}^\top  \\
 -\mathbf{u}_{m_1} & \mathbf{I}_{m_1} & \mathbf{O}_{m_1,m_2} & \dots & \mathbf{O}_{m_1,m_s} \\
 -\mathbf{u}_{m_2} & \mathbf{O}_{m_2,m_1} & \mathbf{I}_{m_2} & \dots & \mathbf{O}_{m_2,m_s} \\
  \vdots & \vdots & \ddots & \ddots & \vdots \\
  -\mathbf{u}_{m_s} & \mathbf{O}_{m_s,m_1} & \dots & \mathbf{O}_{m_s,m_{s-1}} & \mathbf{I}_{m_s}
 \end{matrix}\right)\\
 &=&\left(
\begin{matrix}
 0 & \vec{0}_{m_1}^\top & \vec{0}_{m_2}^\top & \dots & \vec{0}_{m_s}^\top  \\
-\mathbf{B}_{m_1}\mathbf{u}_{m_1} & \mathbf{B}_{m_1} & \mathbf{O}_{m_1,m_2} & \dots & \mathbf{O}_{m_1,m_s} \\
-\mathbf{B}_{m_2}\mathbf{u}_{m_2} & \mathbf{O}_{m_2,m_1} & \mathbf{B}_{m_2} & \dots & \mathbf{O}_{m_2,m_s} \\
  \vdots & \vdots & \ddots & \ddots & \vdots \\
-\mathbf{B}_{m_s}\mathbf{u}_{m_s} & \mathbf{O}_{m_s,m_1} & \dots & \mathbf{O}_{m_s,m_{s-1}} & \mathbf{B}_{m_s}
 \end{matrix}\right)\, .
\end{eqnarray*}
The diagonal elements are clearly non-positive being $0$ and the diagonal elements of the matrices $\mathbf{B}_{m_j}$, $j=1,\dots,n$, namely
\begin{equation}
\label{eq:Ldiag}
\mathrm{diag}(\mathbf{L})=(0,\Lambda^{(1)},\dots,\Lambda^{(1)},\dots,\Lambda^{(s)},\dots,\Lambda^{(s)})\, .
\end{equation}
Let us now consider the out of diagonal elements. It is straightforward to realize that for any $i\geq 2$ we have $L_{ij}\geq 0$, indeed those terms are associated to the null matrices $\mathbf{O}_{m,m'}$ or to the nilpotent ones, $\mathbf{N}_{m}$, hence those entries are $0$ or $1$. It remains to check the sign of the first column, $L_{1j}$. Those elements are of the form $-\mathbf{B}_{m_k}\mathbf{u}_{m_k}$ for some $k=1,\dots,s$. Because of the definition of the Jordan block~\eqref{eq:Bj} we have
\begin{equation*}
 -\mathbf{B}_{m_k}\mathbf{u}_{m_k} = -\left(\Lambda^{(k)}\mathbf{I}_{m_k}+\mathbf{N}_{m_k}\right)\mathbf{u}_{m_k}=-\Lambda^{(k)}\mathbf{u}_{m_k}-\mathbf{N}_{m_k}\mathbf{u}_{m_k}=(-\Lambda^{(k)}-1,\dots,-\Lambda^{(k)}-1,-\Lambda^{(k)})^\top\, ,
\end{equation*}
that are positive by assuming $\Lambda^{(k)}<-1$.

Let us observe that this last constraint can be relaxed by considering the nilpotent matrix
\begin{equation*}
 \mathbf{N}_{m_j}=\left(
\begin{matrix}
 0 & a & 0 & \dots &0\\
 0 & 0 & a & \dots&0\\
 \vdots & \vdots & \ddots & \ddots & \vdots\\
 0 & 0 & 0 & \ddots &a\\
 0 & 0 & 0 & 0 &0
\end{matrix}\right)\, ,
\end{equation*}
for some $a>0$; indeed the condition for non-negativity of the entries becomes $\Lambda^{(k)}<-a$, and taking $a$ close to $0$ we can relax the latter constraint.

Let us conclude this section by discussing the initial assumption about $\Lambda^{(0)}=0$ to be simple. The proposed method allows to create networks with a prescribed real spectrum and multiplicity, but with one node, say the number $1$, with zero in-degree and maximal out-degree, that is  $k^{(in)}_1=0$ and $k^{(out)}_1=n$ (see Fig.~\ref{fig:figuremi} for an example with $m_0=1$, $m_1=2$, $m_2=3$ and $m_3=1$). From the latter figure, one can also observe the interesting structures induced by the multiplicity, for each $m_j>1$ there is a sort of ``folding fan'' with $m_j$ directed ``sticks'' pointing from the node number $1$ to other nodes connected among them with a directed path. For instance one can observe on the right part the ``folding fan'' $1\rightarrow 3\rightarrow 2 \leftarrow 1$ associated to $m_1=2$, on the bottom part the ``folding fan'' $1\rightarrow 6\rightarrow 5, 1\rightarrow 5 \rightarrow4 \leftarrow 1$ associated to $m_2=3$. Notice that for $m_j=1$ the folding fan ``collapses'' into a directed link, here $1\rightarrow 7$ associated to $m_3=1$. To the best of our knowledge, the relations between the algebraic multiplicity of the Laplace eigenvalues and the above topological network motifs is new and deserves to be further investigated in the future.
\begin{figure}[ht]
\centering
\includegraphics[scale=0.45]{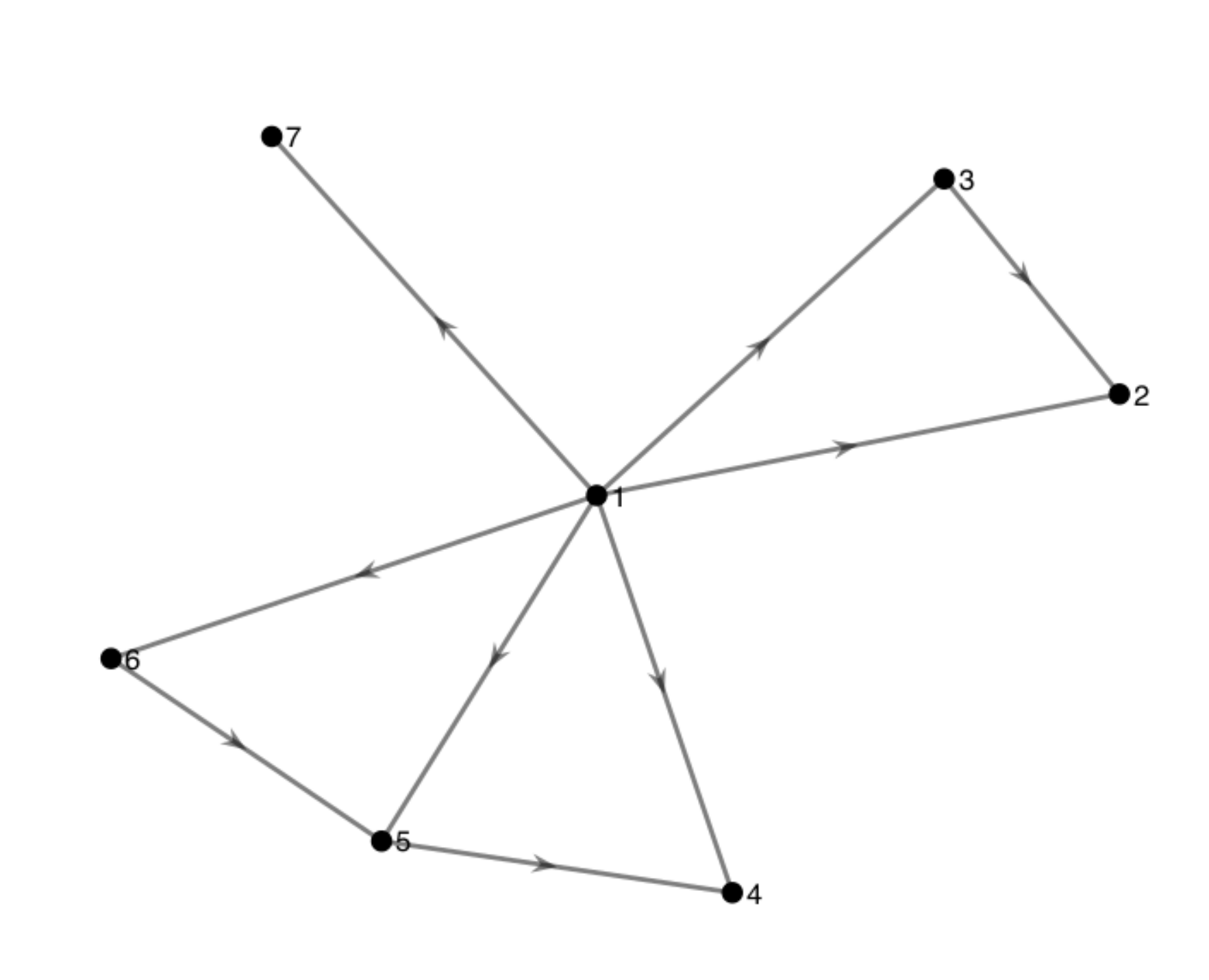}
%\vspace{-2.5cm}
\caption{An example of directed defective network obtained with the previous algorithm by using the multiplicity : $m_0=1$, $m_1=2$, $m_2=3$ and $m_3=1$.}
\label{fig:figuremi}
\end{figure} 

On the other hand, one can obtain a larger network by ``gluing'' together several networks built with the proposed method, seen thus as basic building blocks. Moreover by carefully choosing where to add the new glue-links we can create a network with the eigenvalue $0$ with higher algebraic multiplicity (see Fig.~\ref{fig:figureblocks2}). In the left panel of the Figure we show two basic building blocks, the rightmost one, $\mathcal{N}_1$ made by $5$ nodes, with $(m_0,m_1,m_2,m_3)=(1,1,2,1)$  and $\Lambda_{\mathcal{N}_1}\sim \left(0, -5.90, -7.25, -8.27\right)^\top$, and the rightmost one, $\mathcal{N}_2$ containing $8$ nodes, with $(m_0,m_1,m_2,m_3)=(1,1,1,2,3)$ and $\Lambda_{\mathcal{N}_2}\sim \left(0, -1.36, -3.36, -6.74, -9.33\right)^\top$. On the right panel we show the network, $\mathcal{N}$ with $16$ nodes, obtained by adding the glue-link $7\rightarrow 4$. One can observe that the latter network contains two nodes, $1$ and $6$, with $k_i^{(in)}=0$ and can thus show that this implies that $\Lambda^{(0)}=0$ has now algebraic multiplicity $2$, indeed we can compute $\Lambda_{\mathcal{N}}\sim \left(0, -1.36, -3.36, -6.74, -6.90,-7.25,-8.27,-9.33\right)^\top$ with multiplicity $m_{\mathcal{N}}=\left(2, 1 ,1, 2 ,1,2,1,3\right)^\top$. Let us finally observe that the spectrum of $\Lambda_{\mathcal{N}}$ is almost the union of the spectra of $\Lambda_{\mathcal{N}_1}$ and $\Lambda_{\mathcal{N}_2}$, the only difference being the eigenvalue $- 6.90$ replacing $- 5.90$, this is because the added glue-link modified the weighted in-degree of node $4$ by adding a new weight $1$.
\begin{figure}[ht]
\centering
\includegraphics[scale=0.32]{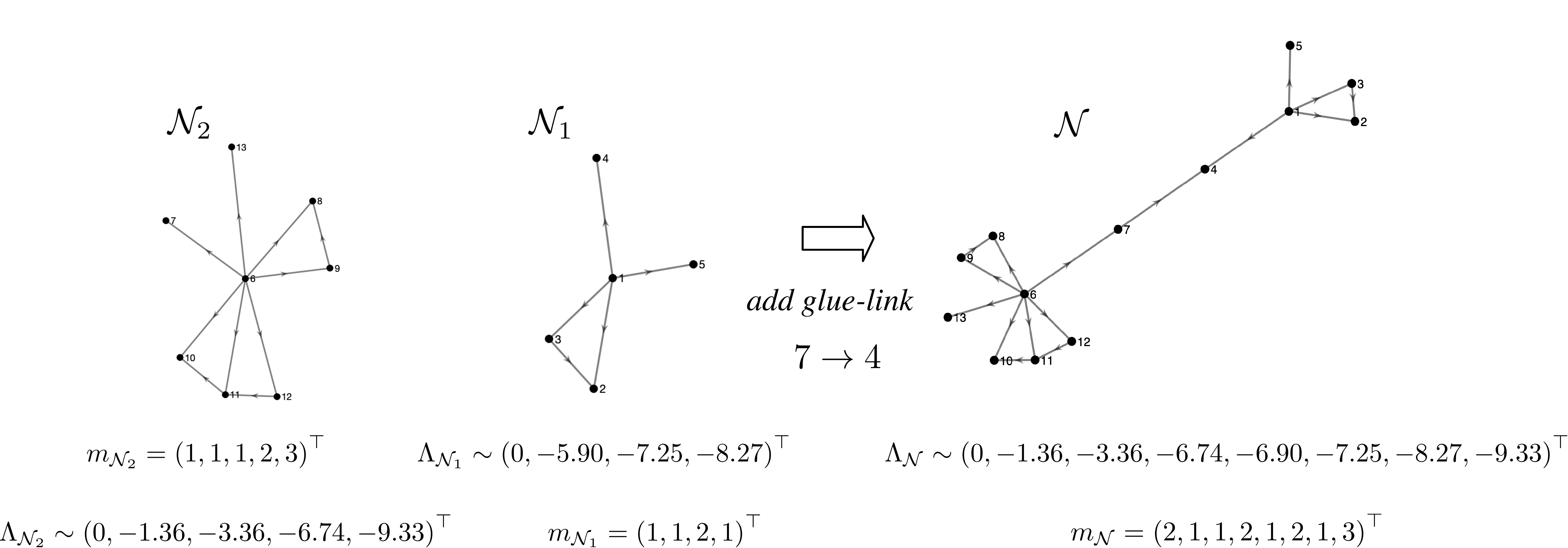}
%\vspace{-2.0cm}
\caption{Creating a larger network by gluing basic building blocks. The obtained network exhibits a $0$ eigenvalue with algebraic multiplicity $2$.}
\label{fig:figureblocks2}
\end{figure} 

Let us observe that by using the idea of gluing together basic building blocks, we can obtain networks that  preserve the degenerate eigenvalues, their multiplicity and only slightly change the remaining eigenvalues, hence without modifying the multiplicity of $0$, an example is shown in Fig.~\ref{fig:figureblocks}. On the left panel we present two basic building blocks similar to the ones used in the construction presented in Fig.~\ref{fig:figureblocks2}, i.e., same multiplicity but different eigenvalues, while on the right part we show the network obtained by adding the glue-link $7\rightarrow 1$, say $\tilde{\mathcal{N}}$. 

The new network has a Laplace matrix whose spectrum is the union of the spectra of the Laplace matrices for the two smaller networks together with a new eigenvalue $-1$, in particular the degenerate eigenvalues did not change their values neither their multiplicity. The reason being that the glue-link $7\rightarrow 1$ will not modify the in-degree of node $7$, or of any other node, but the one of node $1$, the latter initially was $0$ and now becomes $1$. Algebraically the adjacency matrix of $\tilde{\mathcal{N}}$ is given by
\begin{equation*}
 \tilde{\mathbf{A}}=\left(
\begin{matrix}
 \mathbf{A}^{(1)} & \mathbf{E}^{(17)}\\
 \mathbf{O}_{5,8} & \mathbf{A}^{(2)}
\end{matrix}
\right)
\end{equation*}
where the matrix $\mathbf{E}^{(17)}$ has all zero entries but $E^{(17)}_{17}=1$, and $\mathbf{A}^{(j)}$ denotes the adjacency matrix of the network $\tilde{\mathcal{N}}_j$, $j=1,2$. Hence the Laplace matrix of $\tilde{\mathcal{N}}$ results to be
\begin{equation*}
 \tilde{\mathbf{L}}=\left(
\begin{matrix}
\hat{\mathbf{L}}^{(1)} & \mathbf{E}^{(17)}\\
 \mathbf{O}_{5,8} & \tilde{\mathbf{L}}^{(2)}
\end{matrix}
\right)
\end{equation*}
where $\hat{\mathbf{L}}^{(1)}= \tilde{\mathbf{L}}^{(1)}- \mathbf{E}^{(11)}$, namely there is a ``$-1$'' in position $11$, while $\tilde{L}^{(1)}_{11}=0$ (node $1$ has in-degree $0$ in the network $\mathcal{\mathcal{N}}_1$). Because the spectra of $\hat{\mathbf{L}}^{(1)}$ and $\tilde{\mathbf{L}}^{(2)}$ are disjoint, the Jordan Canonical Form of $\tilde{\mathbf{L}}$ is the direct sum of the Jordan Canonical Form of $\hat{\mathbf{L}}^{(1)}$ and $\tilde{\mathbf{L}}^{(2)}$, from which the conclusion about the spectra and the multiplicity easily follows.
\begin{figure}[ht]
\centering
\includegraphics[scale=0.3]{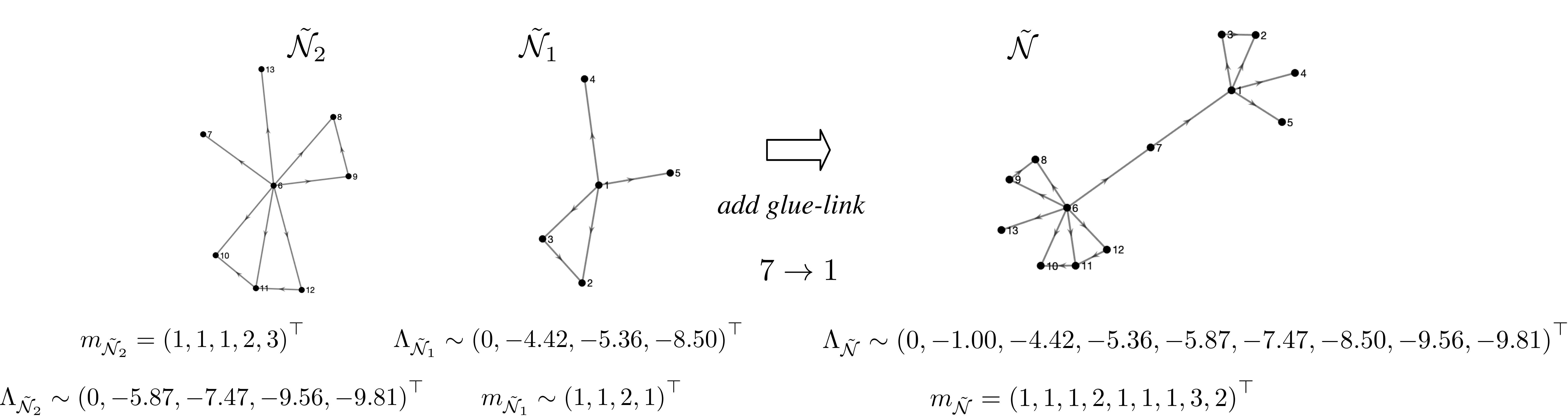}
%\vspace{-2.0cm}
\caption{Creating a larger network by gluing basic building blocks. The spectrum of the obtained network exhibits the same degenerate eigenvalues, multiplicity and only slightly change the remaining eigenvalues with respect to the spectra of the basic building blocks.}
\label{fig:figureblocks}
\end{figure} 
}}

\section{Robustness of pattern reconstruction}
\label{eq:otherex}

The aim of this Section is to present some results supporting the claim that the proposed method for pattern reconstruction based on the use of generalized eigenvectors, is robust with respect to the number of involved (generalized) unstable eigenvectors and the time at which the pattern is considered. \textcolor{black}{In a second moment we will also study the impact of the network size on the reconstruction error.}

\begin{figure}[h]
    \centering
\includegraphics[scale=0.35]{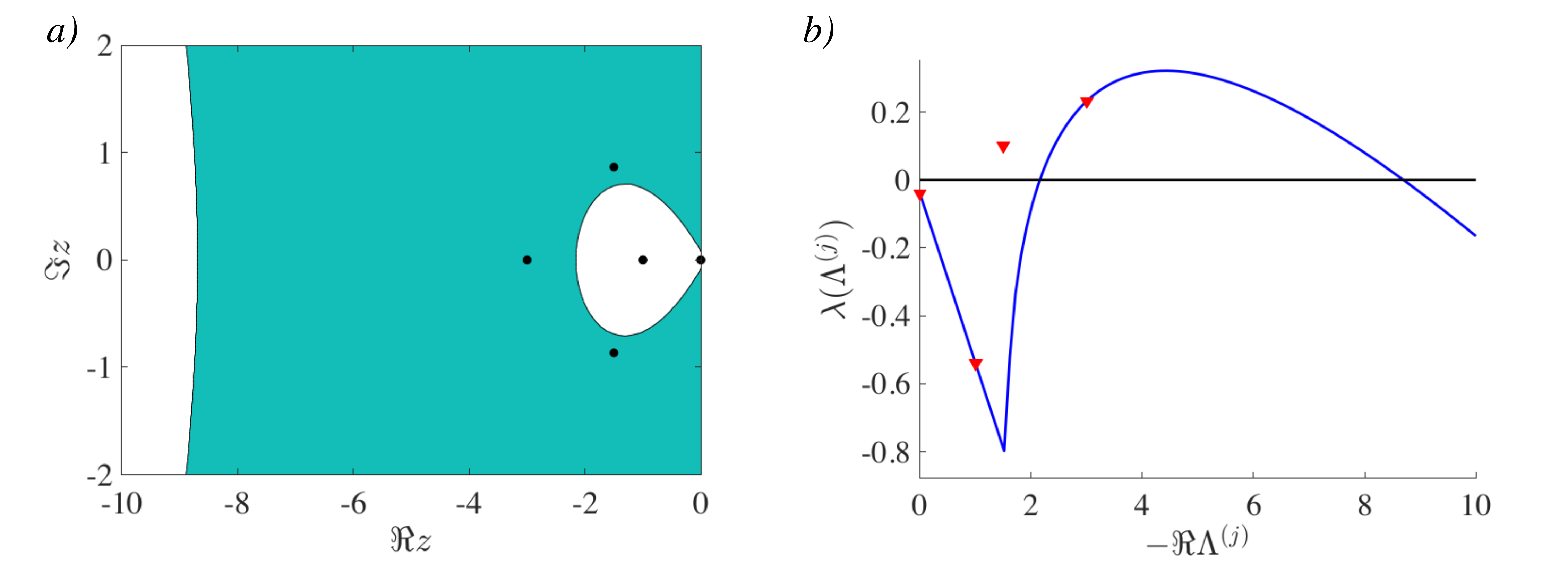}
    \caption{Region of the complex plane associated to Turing instability (panel a)) and dispersion relation (panel b)) computed for the Brusselator model with parameters $b=3.92$, $c=3$, $D_u=0.2$ and $D_v=0.8$. The black dots denote the eigenvalues of the Laplace matrix, $\Lambda^{(1)}=0$ (multiplicity $3$), $\Lambda^{(2)}=-1$ (multiplicity $3$), $\Lambda^{(3)}=-3$ (multiplicity $2$), $\Lambda^{(4)}=-1.5+i 0.866$ (multiplicity $1$) and  $\Lambda^{(5)}=-1.5-\mathrm{i} 0.866$ (multiplicity $1$).}
    \label{fig:ReginstApp}
\end{figure}

\begin{figure}[h]
    \centering
    \includegraphics[scale=0.25]{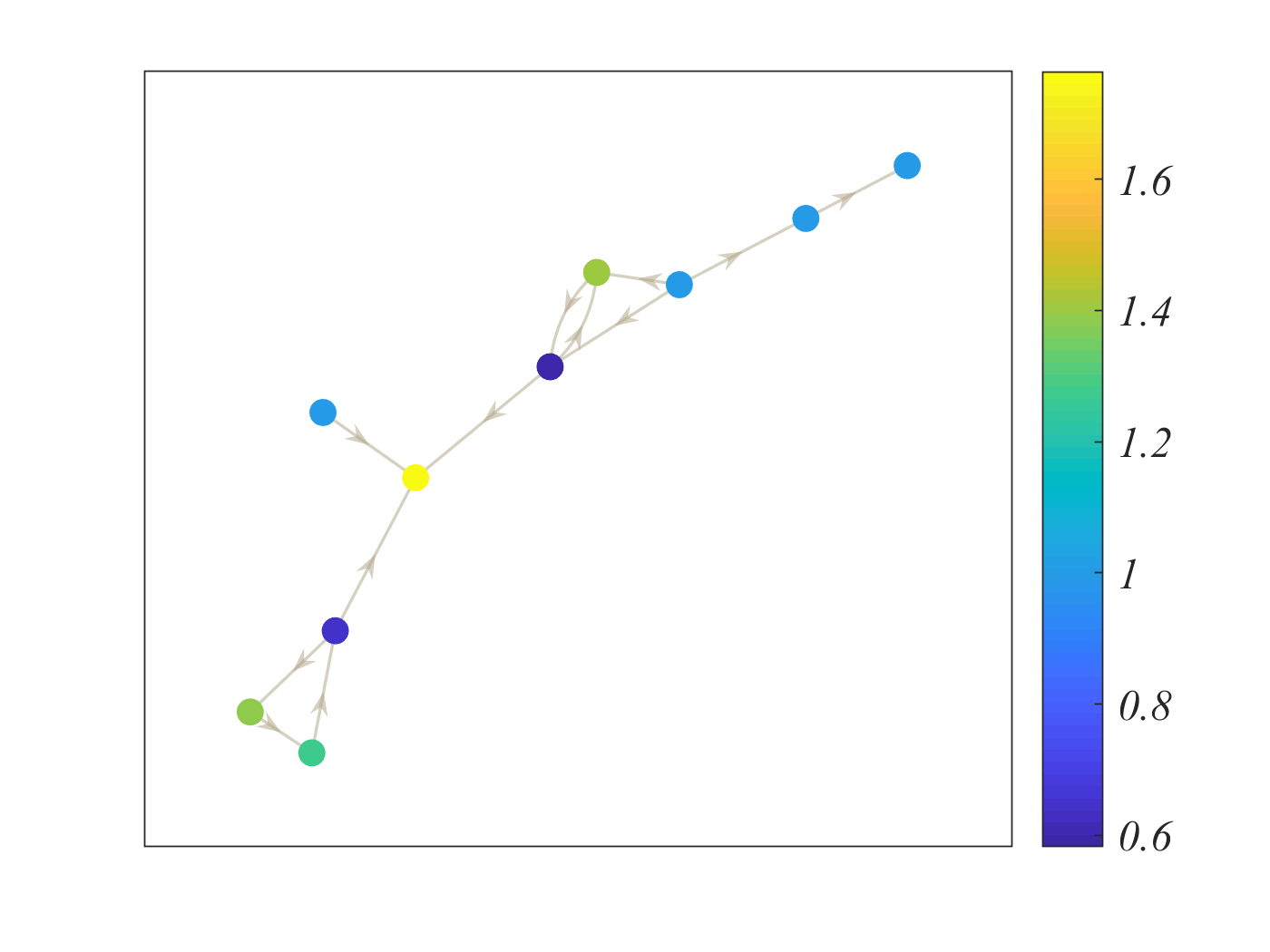}
    \caption{Random non-normal defective network composed by $n=10$ nodes, built by using a directed Erd\H{o}s-R\'enyi where the probability to create a bidirectional link is $0.2$ and the probability to transform it into a directed one is $0.6$. \textcolor{black}{Each node has been shown by using a color code corresponding to the concentration of species $u$ at time $\hat{t}=200$ (see color map).}}
    \label{fig:3 unstable eigenvalue - networkApp}
\end{figure}

Let us first consider a case where there are two unstable eigenvectors and two unstable generalized eigenvectors. In the left panel of Fig.~\ref{fig:ReginstApp} we show the region of instability for the Brusselator model defined on a defective network composed by $N=10$ nodes (see Fig.~\ref{fig:3 unstable eigenvalue - networkApp}), where $\Lambda^{(1)}=0$ and $\Lambda^{(2)}=-1$ both have multiplicity $3$ and are stable, thus they do not intervene in the Turing instability, while $\Lambda^{(3)}=-3$ has multiplicity $2$ and is unstable as well as the complex eigenvalue $\Lambda^{(4)}=-1.5+\mathrm{i} 0.866$ and its complex conjugated $\Lambda^{(5)}=-1.5-\mathrm{i} 0.866$, both with multiplicity one.

\begin{figure}[ht!]
    \centering
   \includegraphics[width=\linewidth]{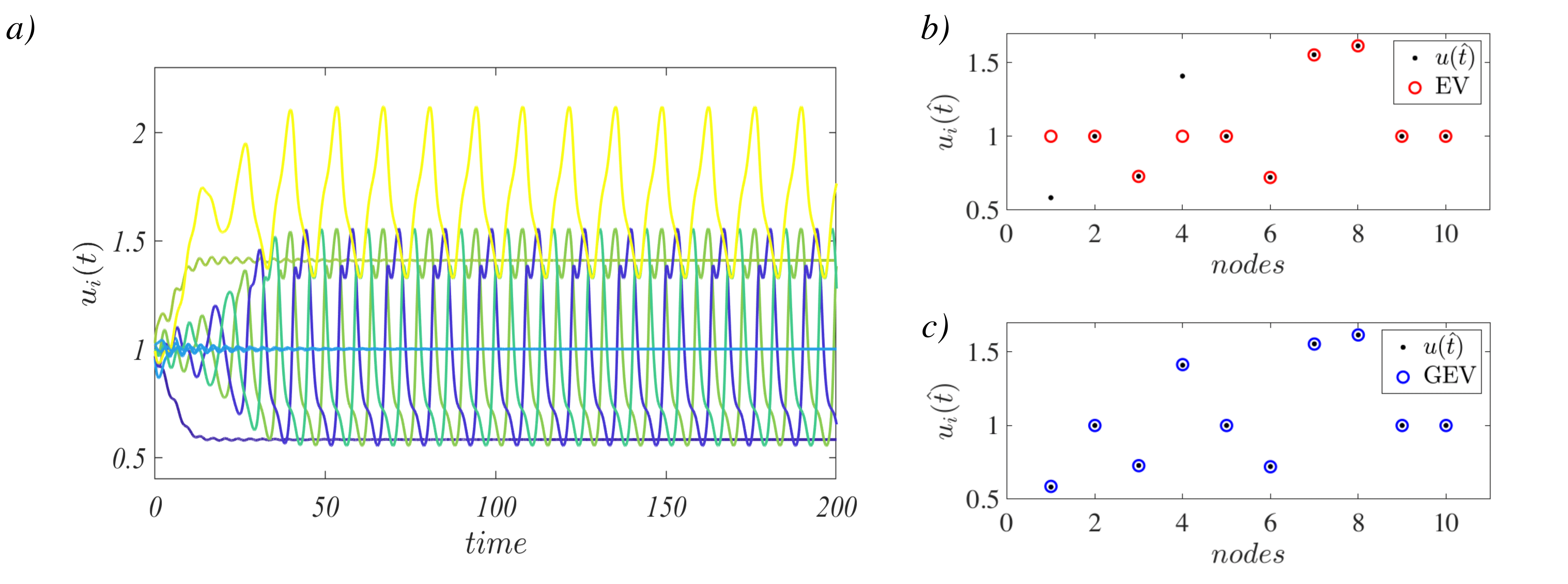}
    \caption{Evolution of the concentration of specie $u_i$ over time (panel a)) for the Brusselator model and pattern reconstruction by using only unstable eigenvectors (panel b)) and unstable eigenvectors and generalized ones (panel c)). 
    In the former case, the reconstruction error is $\varepsilon=0.0623$ while in the latter we have $\varepsilon=0.0012$.
    The model parameters are given by $b=3.92$, $c=3$, $D_u=0.2$ and $D_v=0.8$. The underlying network is the one shown in Fig.~\ref{fig:3 unstable eigenvalue - networkApp}. \textcolor{black}{Each orbit has been shown by using a color code corresponding to the concentration of species $u$ at time $\hat{t}=200$, the same color used in Fig.~\ref{fig:3 unstable eigenvalue - networkApp}.}}
    \label{fig:3 unstable eigenvalue - patternApp}
\end{figure}

In the right panel of Fig.~\ref{fig:ReginstApp} we report the dispersion relation computed for the Brusselator defined on the defective network shown in Fig.~\ref{fig:3 unstable eigenvalue - networkApp}. The Laplace eigenvalues are represented by symbols (red triangles) while the continuous curve is the dispersion relation computed for the $1$-(complex) parameter family of linear systems introduced above with the matrix $\mathbf{M}_j$. One can observe that $\lambda(\Lambda^{(j)})$ is positive for $\Lambda^{(3)}$ and $\Lambda^{(4)}$ and thus the equilibrium $(u_*,v_*)\sim (1,1.3067)$ is unstable, as confirmed from the results shown in panel a) of Fig.~\ref{fig:3 unstable eigenvalue - patternApp} where we report $u_i(t)$ vs. time. Observe that the same conclusion can be obtained by looking at the left panel of Fig.~\ref{fig:ReginstApp} where we can realize that $\Lambda^{(3)}$ and $\Lambda^{(4)}$ lie inside the instability region.
In the right panel of Fig.~\ref{fig:3 unstable eigenvalue - patternApp} we represent the pattern reconstruction at a given fixed $\hat{t}$ both with only unstable eigenvectors and with generalized ones. In the first case the reconstruction error is given by $\varepsilon=0.0623$ while in the second we have $\varepsilon=0.0012$, let us observe that the former is larger even if the factor accounting for the number of used vectors is smaller than one; indeed $d/N_u=3/4$, because we used the eigenvector associated to $\Lambda^{(3)}=-3$ and the two real vectors obtained from the complex eigenvectors associated to $\Lambda^{(4)}=-1.5+i 0.866$. We can then again conclude that the reconstructed pattern obtained by using both eigenvectors and generalized ones has a smaller error than the pattern obtained with only the eigenvectors.

\begin{figure}[ht!]
    \centering
    \includegraphics[scale=0.45]{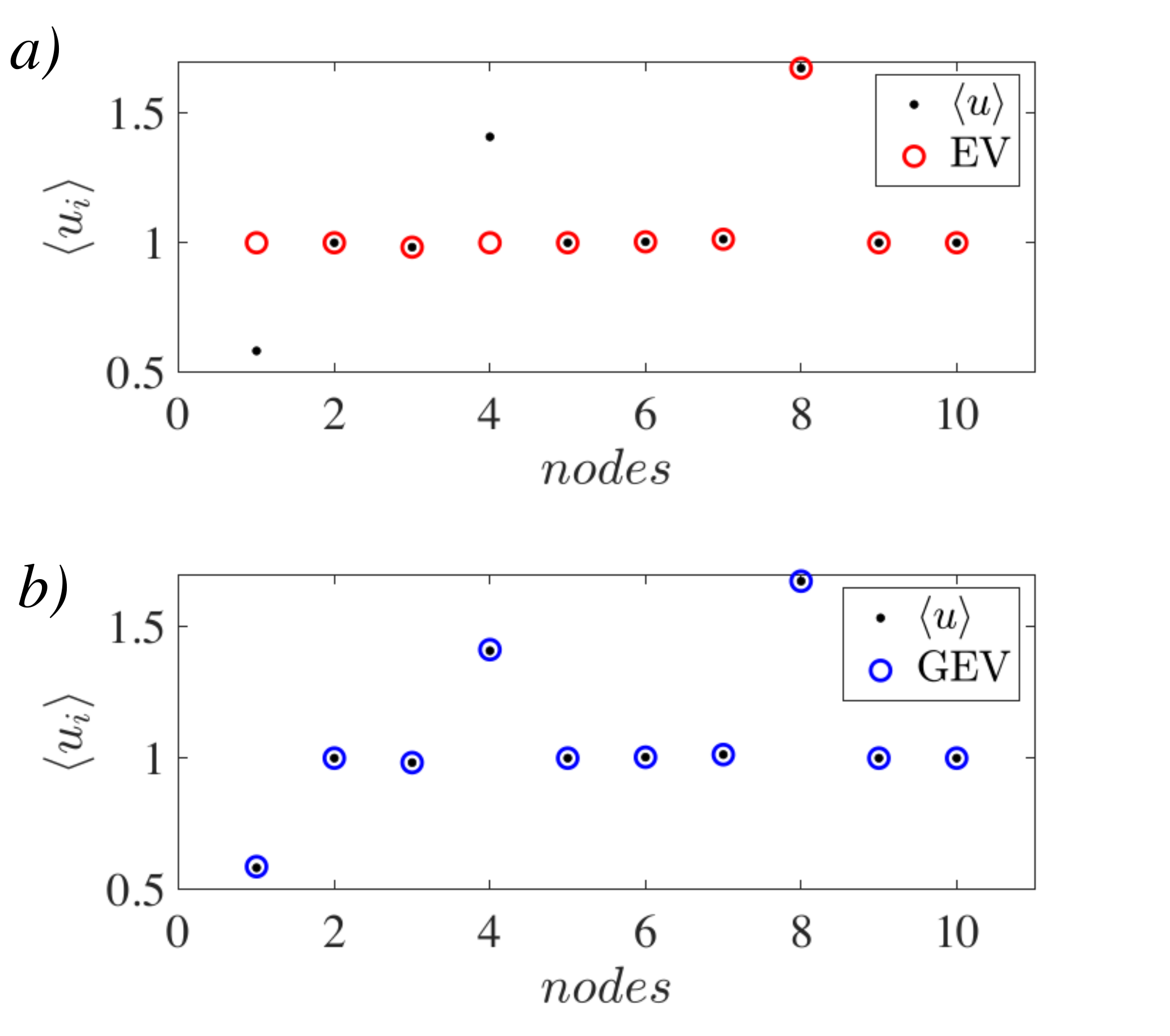}
    \caption{Time average pattern $\langle u \rangle$ vs. reconstructed pattern for the Brusselator model with parameters $b=3.92$, $c=3$, $D_u=0.2$ and $D_v=0.8$. Panel a): the reconstruction is obtained with only the eigenvectors (EV). Panel b): the eigenvectors and generalized eigenvectors (GEV) are used for the reconstruction. In the former case the error is $\varepsilon=0.0620$ while in the latter we get $\varepsilon=8.5\, 10^{-4}$.}
    \label{fig:3 unstable eigenvalue - mean pattern reconstruction}
\end{figure}

In the last example, the concentrations $u_i(t)$ oscillate in time, it would thus be interesting to reconstruct the time average pattern
\begin{equation*}
\langle u_i \rangle=\frac{1}{T}\int_{t_0}^{t_0+T}u_i(s)\,ds \, ,
\end{equation*}
where $t_0>0$ is a sufficiently large lag of time needed to remove a transient phase in the orbit behavior, and $T$ is a (multiple of the) orbit period. Indeed we can show that the proposed scheme works equally well if we want to reconstruct the time average pattern, as shown in Fig.~\ref{fig:3 unstable eigenvalue - mean pattern reconstruction}, where we report the time average patterns and their reconstruction by using the unstable eigenvectors (top panel) and the unstable eigenvectors and generalized ones for the same networks and model parameters used for Fig.~\ref{fig:3 unstable eigenvalue - patternApp}. In the former case the reconstruction error is $\varepsilon=0.0620$ while in the second one we obtain $\varepsilon=8.5\, 10^{-4}$.

To conclude this section let us consider a larger network, shown in Fig.~\ref{fig:20 nodes, 3 unstable eigenvalues - network}, made by $20$ nodes and exhibiting three unstable eigenvalues, $\Lambda^{(4)}=-4$ with multiplicity four, $\Lambda^{(5)}=-5$ and $\Lambda^{(6)}=-7$ each with multiplicity one, and three stable ones, $\Lambda^{(3)}=-2$ with multiplicity seven, $\Lambda^{(2)}=-1$ with multiplicity three and $\Lambda^{(1)}=0$ with multiplicity four.
\begin{figure}[ht]
    \centering
    \includegraphics[scale=0.2]{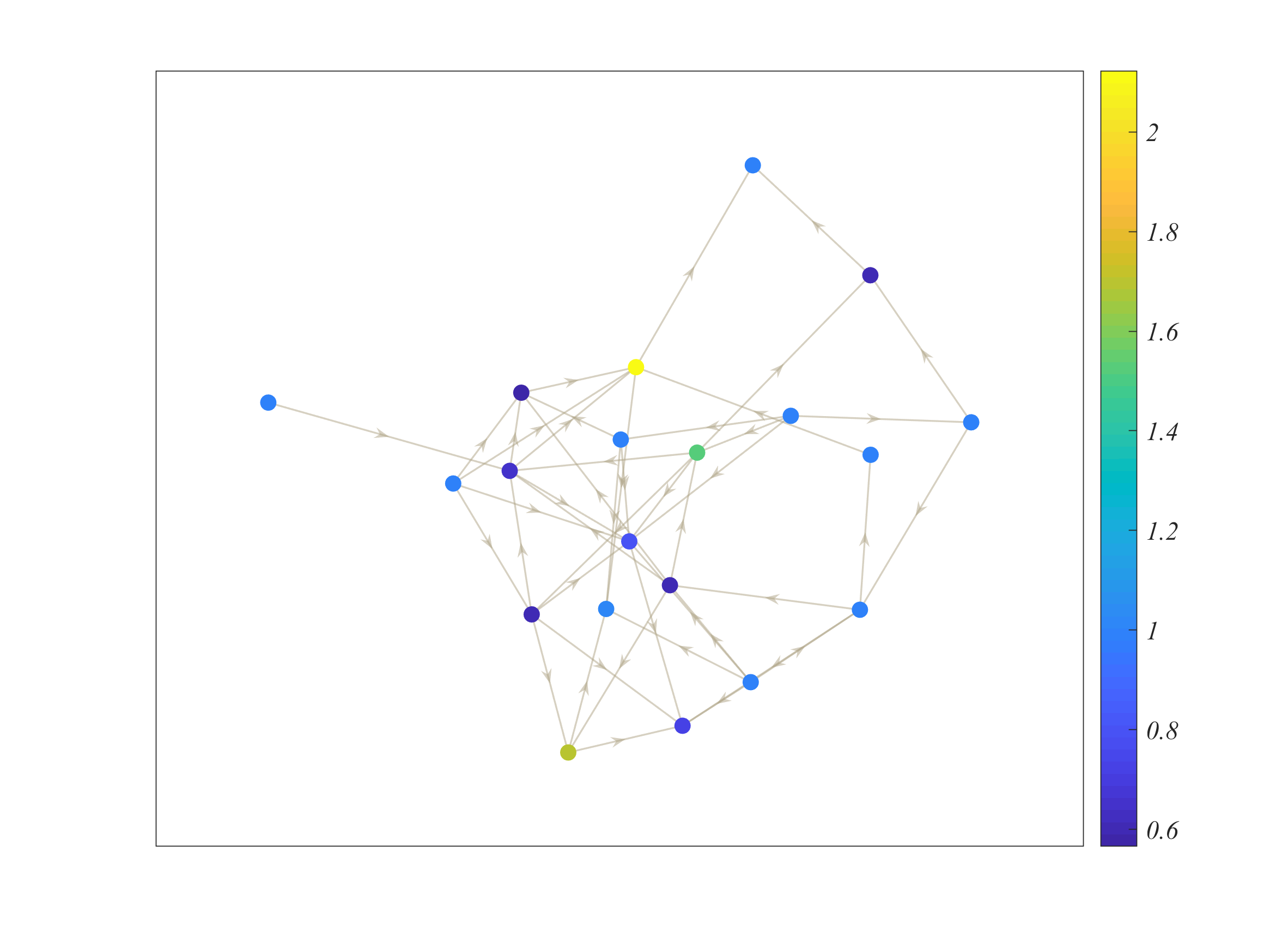}
    \caption{Random non-normal defective network composed by $n=20$ nodes, built by using a directed Erd\H{o}s-R\'enyi where the probability to create a bidirectional link is $0.2$ and the probability to transform it into a directed one is $0.6$. \textcolor{black}{Each node has been shown by using a color code corresponding to the concentration of species $u$ at time $\hat{t}=200$ (see color map).}}
    \label{fig:20 nodes, 3 unstable eigenvalues - network}
\end{figure}

The region of instability shown in panel a) of Fig.~\ref{fig: 20 nodes, 3 unstable eigenvalue - dispersion relation} or equivalently the dispersion relation, panel b) of the same figure, testify the existence of three unstable modes and indeed one can observe the emergence of patterns (see panel a) of Fig.~\ref{fig: 20 nodes, 3 unstable eigenvalue - pattern}). Looking at the dispersion relation one can observe that the latter assumes values very close once evaluated at $\Lambda^{(4)}=-4$ and $\Lambda^{(5)}=-5$, indeed $\lambda(-4)=0.315$ and $\lambda(-5)=0.312$.
\begin{figure}[ht!]
    \centering
    \includegraphics[scale=0.35]{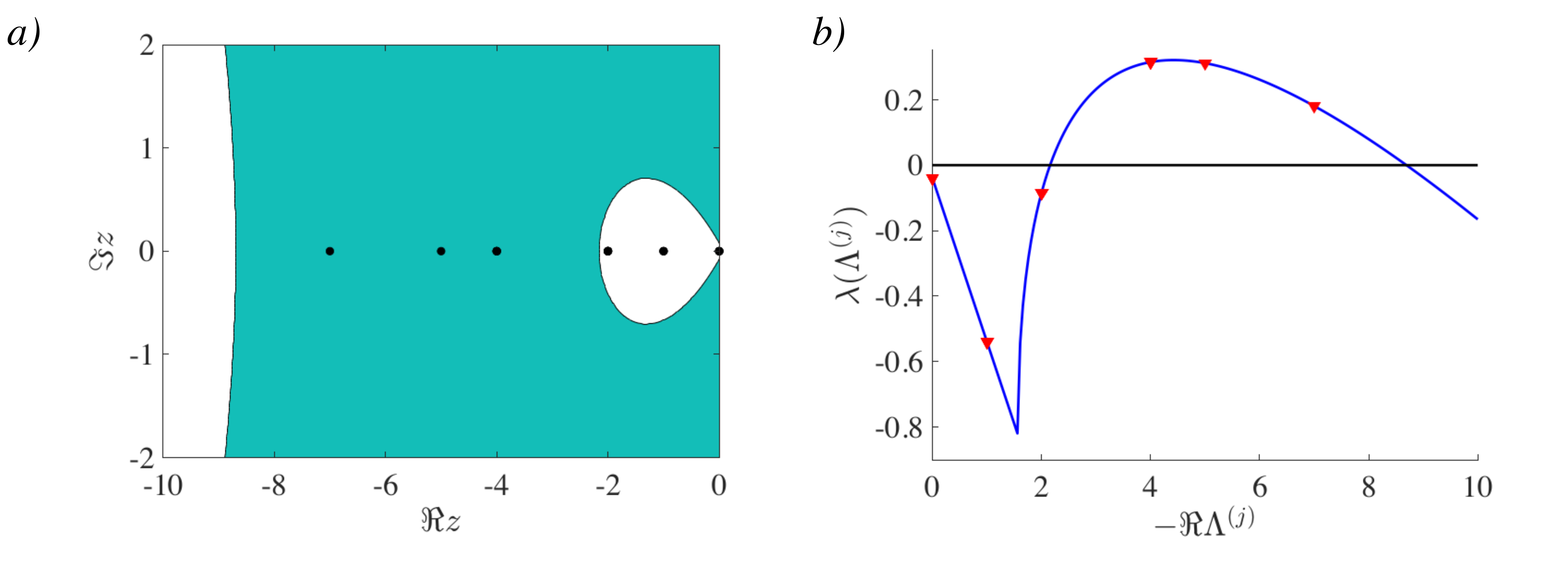}
    \caption{Region of the complex plane associated to Turing instability (panel a)) and dispersion relation (panel b)) computed for the Brusselator model with parameters $b=3.92$, $c=3$, $D_u=0.2$ and $D_v=0.8$. The black dots denote the eigenvalues of the Laplace matrix, $\Lambda^{(1)}=0$ (multiplicity $4$), $\Lambda^{(2)}=-1$ (multiplicity $3$), $\Lambda^{(3)}=-2$ (multiplicity $7$), $\Lambda^{(4)}=-4$ (multiplicity $4$), $\Lambda^{(5)}=-5$ and $\Lambda^{(6)}=-7$ each with multiplicity one. The used network is shown in Fig.~\ref{fig:20 nodes, 3 unstable eigenvalues - network}.}
    \label{fig: 20 nodes, 3 unstable eigenvalue - dispersion relation}
\end{figure}

The most unstable mode drives the onset of the instability, however it is not clear a priori if the pattern would be better reconstructed by using the most unstable mode alone, $\Lambda^{(4)}=-4$, together with the associated generalized eigenvectors or the mode $\Lambda^{(5)}=-5$. By eyeball analysis of panels b)-e) of Fig.~\ref{fig: 20 nodes, 3 unstable eigenvalue - pattern} one would conclude that the strategy relying on the use of $\Lambda^{(4)}=-4$ and the generalized eigenvectors provides the better results. Let us however observe that the smallest reconstruction error is found by using $\Lambda^{(4)}=-4$ alone; this is due to the dimensionality rescaling factor, here $1/4$, corresponding thus to the use of one eigenvector out of four possible (generalized) eigenvectors. This analysis should thus be considered as a preliminary step toward a deeper understanding of the problem, that will be addressed in a forthcoming study.
\begin{figure}[ht]
    \centering
\includegraphics[width=\linewidth]{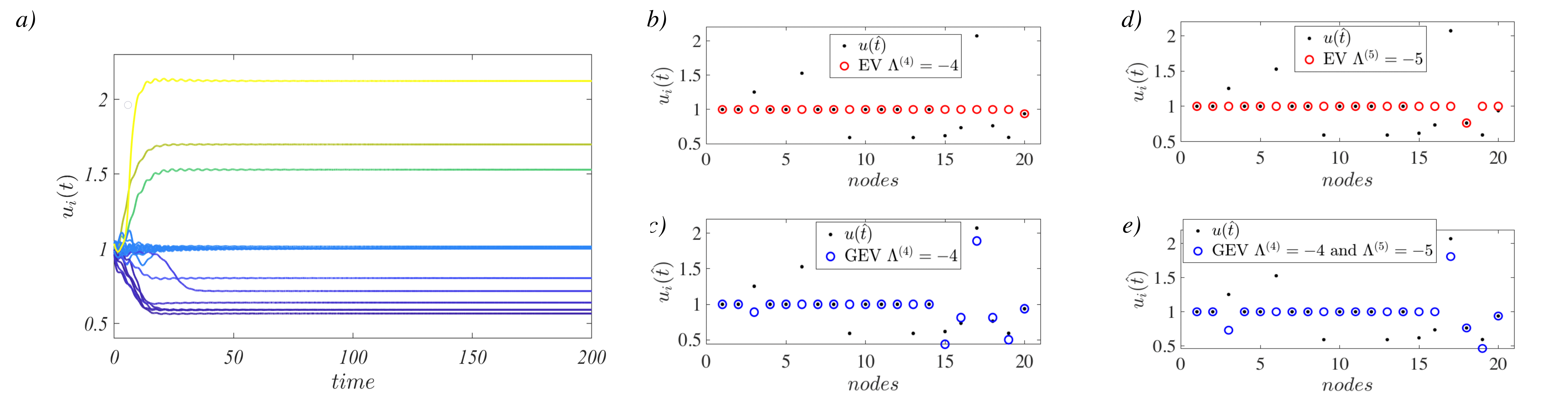}
    \caption{Evolution of the concentration of specie $u_i$ over time (panel a)) for the Brusselator model and pattern reconstruction by using only the unstable eigenvector associated to $\lambda^{(4)}=-4$ (panel b)) and the unstable eigenvector associated to $\lambda^{(5)}=-5$ (panel d)). In panel c) we report the pattern reconstructed by using the unstable eigenvector for $\lambda^{(4)}=-4$ and its associated generalized eigenvectors. Finally in panel e) we report the patterns obtained by using the unstable eigenvectors associated to $\lambda^{(4)}=-4$ and $\lambda^{(5)}=-5$ and the generalized eigenvectors associated to the former vector. The model parameters are given by $b=3.92$, $c=3$, $D_u=0.2$ and $D_v=0.8$. The underlying network is the one show in Fig.~\ref{fig:20 nodes, 3 unstable eigenvalues - network}. \textcolor{black}{Orbits have been reported by using the same color of the corresponding node in Fig.~\ref{fig:20 nodes, 3 unstable eigenvalues - network}, which corresponds to the value of $u$ at $\hat{t}=200$.}}
    \label{fig: 20 nodes, 3 unstable eigenvalue - pattern}
\end{figure}

{\color{black}{We conclude this section by studying the impact of the network size on the error reconstruction. We will use the directed defective networks obtained by using the algorithm presented in Section~\ref{eq:dirdefnet} as support for the Brusselator model to perform this analysis. Moreover, to simplify the analysis and removing for possible confounding factors, we assumed the existence of a unique unstable degenerate eigenvalue while all the remaining stable eigenvalues can be degenerate or not. We then fixed some generic values for the Brusselator model and computed the dispersion relation, namely the largest real part of the eigenvalues of the $1$-(complex) parameter family of the matrix $\mathbf{J}_0 +\zeta \mathbf{D}$, as defined in the main text (see Fig.~\ref{fig:reldispex}).
\begin{figure}[ht]
    \centering
    \includegraphics[scale=0.42]{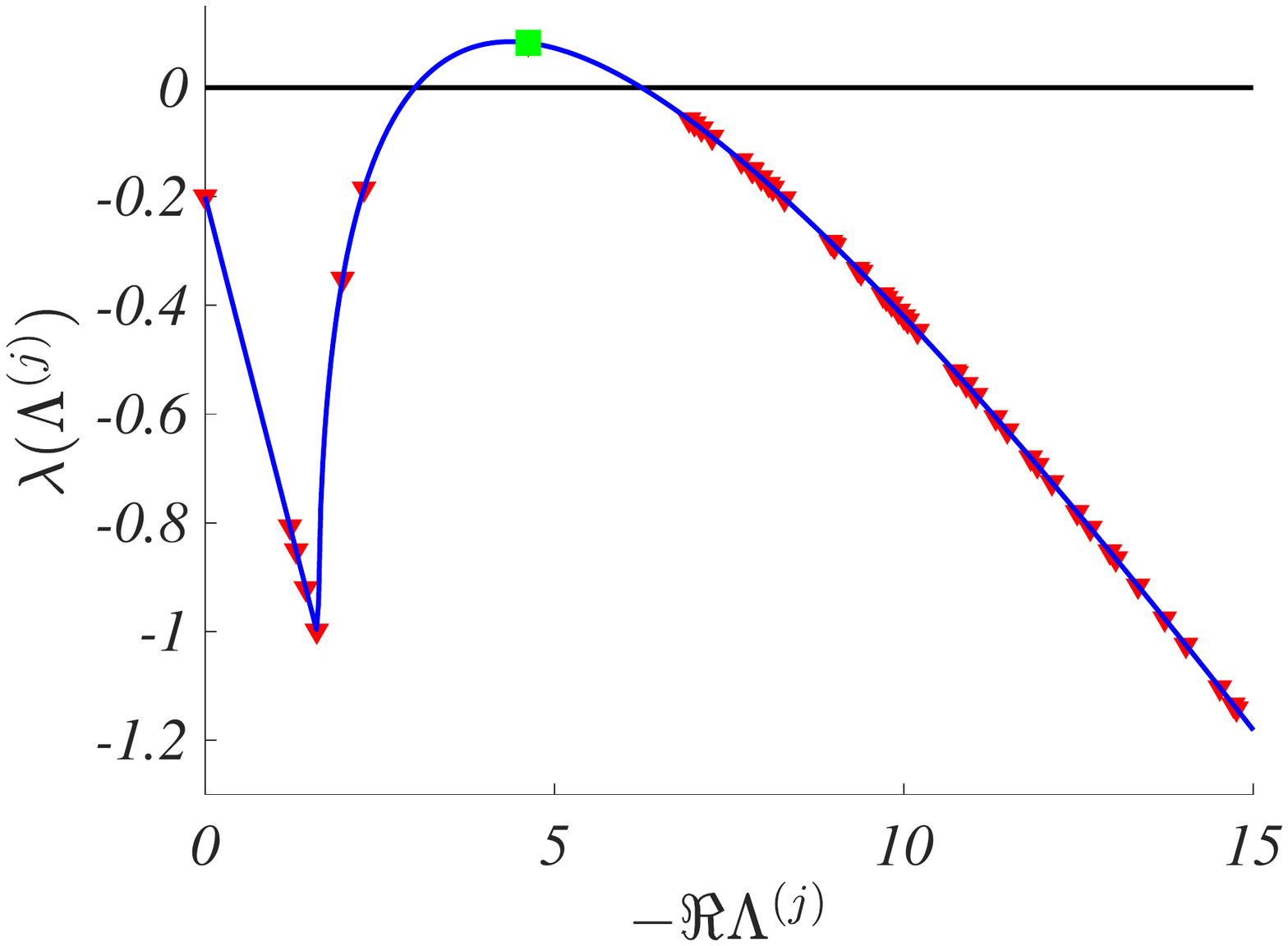}
    \vspace{-2.5cm}
    \caption{Dispersion relation. The largest real part of the spectrum of the matrix $\mathbf{J}_0+\zeta \mathbf{D}$ is shown in blue as a function of $\zeta$, the dispersion relation evaluated on stable Laplace eigenvalues is reported by using red triangles, while the unique unstable eigenvalue is shown with a green square.}
    \label{fig:reldispex}
\end{figure}

The size of the network and the multiplicity are related by $n=\sum_{j=1}^s m_j$. A first study concerns thus the dependence of the pattern reconstruction error~\eqref{eq:werrordef} as a function of the multiplicity, $m$, of the unique unstable eigenvalues for a fixed network size. The results reported in Fig.~\ref{fig:errorvsm} concern a network whose size is $n=1000$ and we let the multiplicity $m$ to vary from $2$ to $20$; for each value of $m$ we compute the pattern reconstruction error by using the unique eigenvector (EV - blue points) and the latter eigenvector together with the $m-1$ generalized eigenvectors (GEV - black squares), let us stress that in the former case the pre-factor ${d}/{N_u}$ in Eq.~\eqref{eq:werrordef} is equal to $1/m$, while it is the unity in the case of generalized eigenvectors. Each point is the average over $100$ independent replicas of the construction, i.e., different networks with a different spectrum but with the same multiplicity. Two main messages can be drawn from these results: first of all, the use of the generalized unstable eigenvectors provides an error or several order of magnitude smaller than the use of the unstable eigenvector. Second, the dependence of the error on the value $m$ is relatively small.
\begin{figure}[ht]
    \centering
    \includegraphics[scale=0.42]{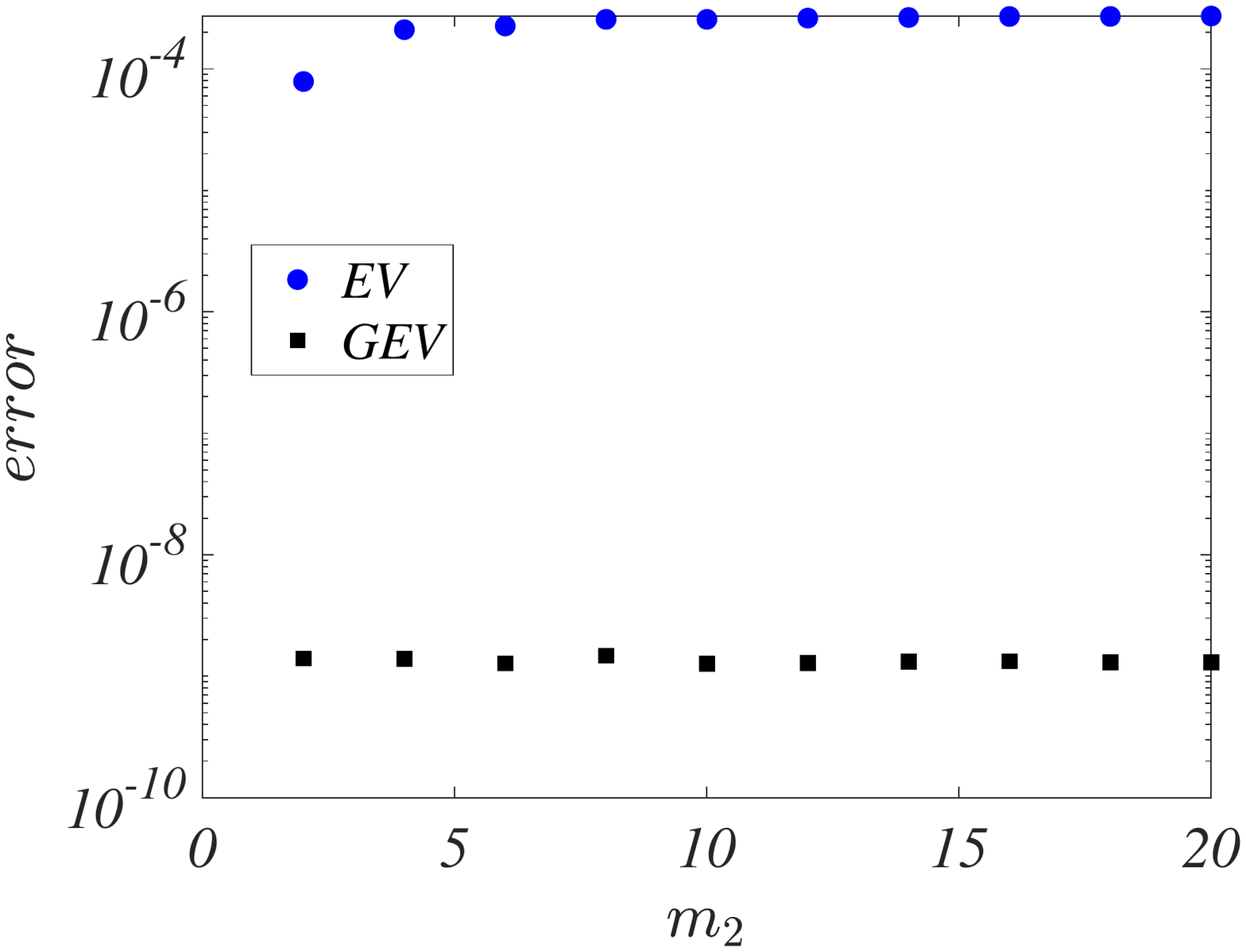}
    \vspace{-2.5cm}
    \caption{The pattern reconstruction error as a function of the multiplicity $m$ of the unique unstable eigenvalues for a network of size $n=1000$.}
    \label{fig:errorvsm}
\end{figure}
 
Because the multiplicity of the unstable eigenvalues does not play a relevant role, we decided to study the error as a function of the network size for randomly generated networks of increasing sizes obtained by applying the gluing construction presented in Section~\ref{eq:dirdefnet}, with the constraint of having a unique unstable eigenvalue, whose geometric multiplicity is a random number uniformly drawn in $\{2,3,4,5\}$. We show in Fig.~\ref{fig:errorvsN} the results for networks whose sizes ranges from $100$ to $1000$ and we can conclude that the error is always smaller in the case of the generalized eigenvectors (GEV (dir) - black squares) are used to reconstruct the pattern with respect to the case where only the unstable eigenvector is used (EV (dir) - blue circles). In the same Figure we report the error computed by using a symmetric network with a single unstable eigenvector built by using the algorithm presented in~\cite{FWD2018} (EV (symm) - red diamonds) and we can observe that the error lies in between the previous two cases.
 \begin{figure}[ht]
    \centering
    \includegraphics[scale=0.42]{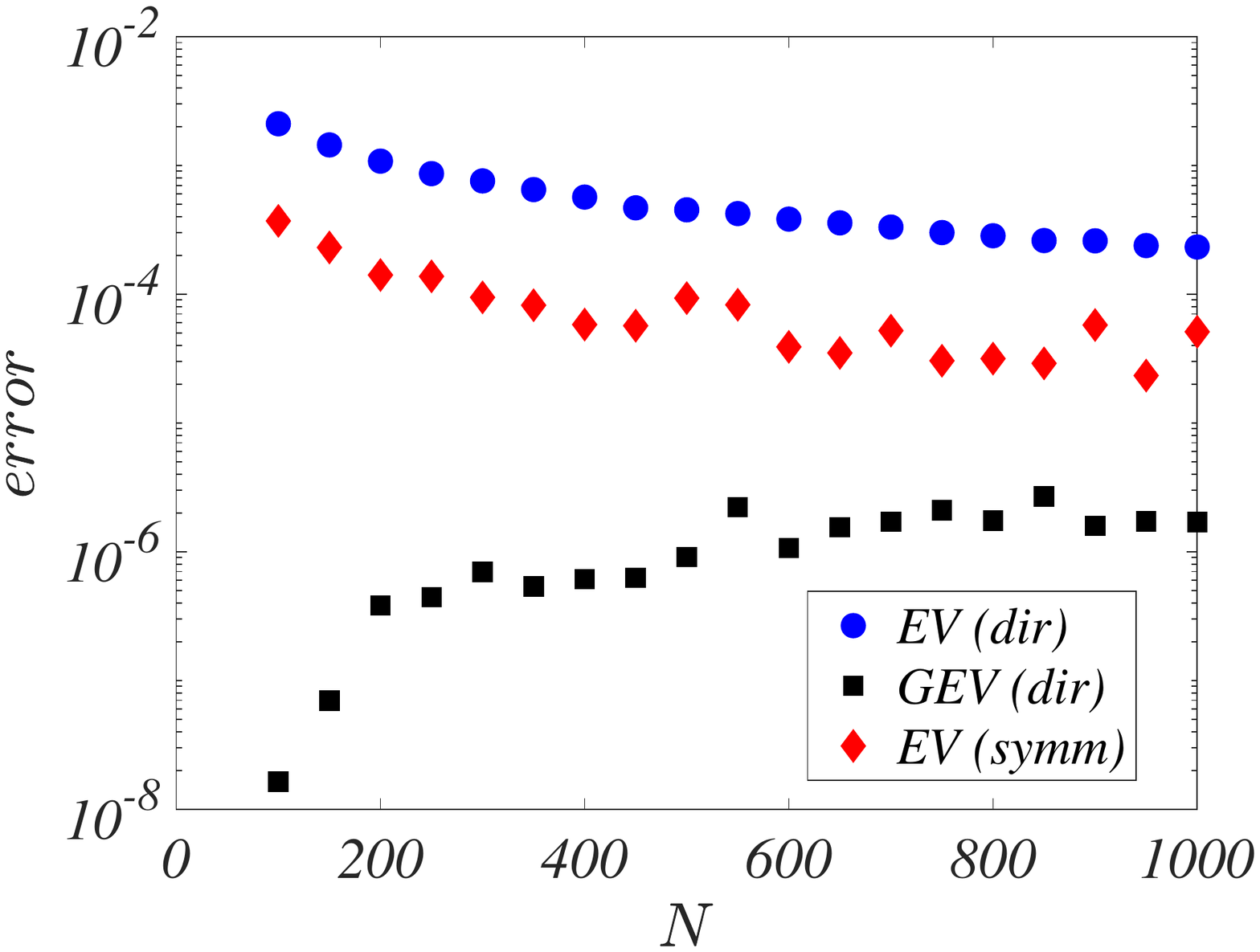}
    \vspace{-2.5cm}
    \caption{The pattern reconstruction error as a function of the network size. In the case of symmetric networks ({\it EV symmetric}, red diamonds) only the unstable eigenvector has been used to reconstruct the pattern; in the case of directed defective networks we can use the unstable eigenvector alone ({\it EV dir}, blue circles) or together with the generalized eigenvectors ({\it GEV dir}, black squares). Each symbols is the result of the average over $100$ independent networks reconstructions.}
    \label{fig:errorvsN}
\end{figure}

We conclude this analysis with some cautionary remarks: the family of directed defective networks we obtain with the above presented algorithm could not be the most general one, in particular because we allow only for real spectra and well localized eigenvectors as in the algorithm proposed in~\cite{FWD2018}, and those facts could induce an unwanted and uncontrolled bias. 
}}
\end{document}